\documentclass[article,moreauthors,pdftex,10pt,a4paper]{mdpi-arxiv}

\Title{Graph Approach to Quantum Teleportation Dynamics}

\Author{E. Honrubia$^{\dagger}$ and A. S. Sanz$^{\ddagger}$*\orcidA{}}

\AuthorNames{E. Honrubia and A. S. Sanz}

\address{%
Department of Optics, Faculty of Physical Sciences,
Universidad Complutense de Madrid,\\
Pza.\ Ciencias 1, Ciudad Universitaria, 28040 Madrid, Spain;\\
$^{\dagger}$ efrenhon@ucm.es,
$^{\ddagger}$ a.s.sanz@fis.ucm.es}

\corres{Correspondence: a.s.sanz@fis.ucm.es; Tel.: +34-91-394-4673}

\abstract{
Quantum teleportation plays a key role in modern quantum technologies.
Thus, it is of much interest to generate alternative approaches or representations
aimed at allowing us a better understanding of the physics involved in the process
from different perspectives.
With this purpose, here an approach based on graph theory is introduced and discussed
in the context of some applications.
Its main goal is to provide a fully symbolic framework for quantum teleportation
from a dynamical viewpoint, which makes explicit at each stage of the process how
entanglement and information swap among the qubits involved in it.
In order to construct this dynamical perspective, it has been necessary to
define some auxiliary elements, namely virtual nodes and edges, as well as an additional
notation for nodes describing potential states (against nodes accounting for actual
states).
With these elements, not only the flow of the process can be followed step by step, but
they allow us to establish  a direct correspondence between this graph-based approach
and the usual state vector description.
To show the suitability and versatility of this graph-based approach, several
particular teleportation examples are examined, which include bipartite, tripartite and
tetrapartite maximally entangled states as quantum channels.
From the analysis of these cases, a general protocol is discussed in the case of sharing
a maximally entangled multi-qubit system.
}


\newcommand{\bd}{\begin{displaymath}}
\newcommand{\ed}{\end{displaymath}}
\newcommand{\be}{\begin{equation}}
\newcommand{\ee}{\end{equation}}
\newcommand{\ba}{\begin{eqnarray}}
\newcommand{\ea}{\end{eqnarray}}

\usepackage{xcolor}
\usepackage{soul}

\begin{document}


\section{Introduction}
\label{sec1}

Quantum entanglement is the cornerstone of modern quantum technologies \cite{milburn:PTRSLA:2003}.
Within them, quantum teleportation, the capability to transfer the information encoded in the
quantum state of a physical
system (e.g., an atom or a photon) to another distant physical system, plays an important role,
as its remarkable theoretical and experimental development has shown over the last quarter of a century
\cite{gisin:entropy:2019}.
Since the first protocol introduced by Bennet et al.~\cite{bennet:PRL:1993},
a number of milestones have paved the way for a solid settlement of the technology
based on it.
In 1997, the first experimental evidence was demonstrated by transferring
a single photon state \cite{bouwmeester:nature:1997}.
These former experiments, where the distance mediating the quantum transfer was of
about a few meters, were followed soon afterwards by the first long-distance
experiments, involving several kilometers \cite{zeilinger:nature:2003}.
In this regard, a record was established in 2012, when the experiments were performed
between two laboratories separated a distance of 143~Km \cite{zeilinger:nature:2012}, with the
purpose to demonstrate the feasibility of the phenomenon at distances of the order
of a low Earth orbit and, hence, to implement quantum key distribution protocols at
a global level.
That record was overcome shortly after by transferring a quantum state between
the cities of Delingha and Lijiang, separated by a distance of 1,200~km \cite{JWPan:science:2017},
and then with the aid of the Micius satellite \cite{JWPan:PRL:2017}, setting a new record and
also the basis for global communications.
Furthermore, there are other important proposals or milestones in recent years regarding quantum teleportation,
such as the possibility to transfer internal quantum states between living organisms (e.g.,
bacteria \cite{li:SciBull:2016}), the use of this phenomenon to acquire information from the
inside of black holes \cite{landsman:nature:2019}, or the experimental realization of quantum
teleportation of qutrits \cite{zelinger-JWPan:PRL:2019}.
No doubt, these appealing studies directly point towards the development
of more sophisticated protocols enabling the teleportation of complex quantum objects.

Nevertheless, regardless of the level of complexity and sophistication involved in all
those investigations, the physical essence of the process still remains the same:
an entangled pair of particles is used as the resource necessary to transfer the
information encoded in a quantum
system to another distant system, in a way that somehow remembers the classical momentum transfer in
the so-called Newton's cradle.
In this regard, one wonders whether it is possible to find a conceptual description of
the process, which might help us to better understand it without making emphasis on the
particular physical substrate, as it is usually the case when we consider the standard
quantum formulation.
This is an idea underlying analogous descriptions of quantum information transfer processes,
such as Penrose's graphical notation \cite{penrose:graphnot:1971} or, more recently,
Coecke's quantum picturialism \cite{coecke:contempphys:2010}, closer to logic.
Another route that can be followed is that of topology.
Kauffman and Lomonaco, for instance, have developed an approach to account for quantum
logic gates and entanglement operators \cite{kauffman:NJP:2002,kauffman:NJP:2004}.
Different topological surfaces have also been used by Mironov \cite{mironov:universe:2019,mironov:JHEP:2019}
to deal with the issue of measuring the Von Neumann entropy.

The present work is motivated by the search of an effective pictorial representation for
quantum teleportation in dynamical terms, beyond the more standard state vector views,
which in the end are more closely connected to particular physical substrates (electrons,
photons, vacancies, etc.).
Typically, topological descriptions and descriptors are developed within a static framework,
just as an operational tool to classify quantum states or to obtain entanglment relations
and measures without involving time.
Quantum teleportation, on the other hand, inherently involves evolution in time as the
information conveyed by a given system is transferred via a quantum channel (and entangled
system) somewhere else, even if time-dependence is not explicitly displayed (this happens,
for instance, in the former protocol introduced by Bennet et al.).
Hence, one wonders whether it is possible to describe this flow, which
involves unitary evolution stages with non-unitary measurements by means of a symbolic
language, which captures the essence of the full process and, in turn, maintains a direct
correspondence with the usual state vector description.
As it is shown here, this can be done with the aid of graph theory, and introducing some
auxiliary elements and notation, which confers the approach with enough flexibility to
mimic a dynamical evolution.
This description, moreover, is also general enough as to provide protocols when quantum
channels consist even of large quantum systems \cite{karlsson:PRA:1998}.
In this regard, the relational approach introduced here is intended to build a wider and
and deeper understanding of quantum dynamical processes involving entanglement, such as
quantum teleportation, which might also be easily transferred to other quantum dynamical
problems in a rather simple form despite of an eventual gradual increment of the level
of complexity involved (e.g., a web or lattice of interconnected qubits).

To tackle the issue, a recent work by Quinta and Andr\'e \cite{quinta:PRA:2018} has
provided us the grounds upon which our graph approach has been constructed.
In this work, a relational classification system for $N$-qubit states is provided, which
is based on how many entangled subsystems are involved, in a way similar to the search of
irreducible representations in group theory.
By recasting then the quantum state in the form of an $N$-variable polynomial, they are
able to associate an $N$-component topological link with such a quantum state.
Based on this idea, and taking into account the construction of graphs from the
corresponding polynomials and links, we have developed an intuitive graph approach
to quantum teleportation processes, which include virtual nodes and edges, on the one
hand, and additional labeling for nodes to denote potential states (in contrast to
the usual actual states), on the other hand.
In this regard, this approach goes beyond more standard approaches based on links, which
lack this relational versatility of graphs, necessary to generate the idea of evolution
or flow in time.

The work has been organized as follows.
To be self-contained, some elementary notions from the polynomial and link approach developed by
Quinta and Andr\'e are introduced and discussed in Sec.~\ref{sec2}.
In order to get the flavor of the approach and hence make more evident the transition to the graph
representation, the particular case of a four-qubit state coupled to an auxiliary six-level qudit
is analyzed.
Then, the graph approach here considered to account for entangled systems is introduced and
discussed in Sec.~\ref{sec3}.
The application to quantum teleportation is presented in Sec.~\ref{sec4}, first in the
case of the standard protocol, and then in the general case of the coupling of a qubit with an
$N$-party entangled state.
To conclude, a series of final remarks are summarized in Sec.~\ref{sec5}.


\section{Polynomials, links and entanglement}
\label{sec2}

According to Quinta and Andr\'e \cite{quinta:PRA:2018}, $N$-qubit entangled states can be represented
by polynomial and topological links, where each constituting qubit is associated with a
variable in the polynomial and a ring in the link structure.
Consider the operation of partially tracing over a particular qubit in the $N$-qubit density
matrix that describes such a state.
Within this representation, this is equivalent to:
\begin{itemize}
 \item[(i)] Setting to zero the variable describing the traced qubit in the corresponding
 polynomial.

\vspace{.25cm}

 \item[(ii)] Removing the ring related to this qubit from the full system topological link.
\end{itemize}
The resulting polynomial and link thus describe the reduced density matrix for the remaining
$N-1$ qubits.
Accordingly, because the subsequent sub-polynomials and sub-links that arise after
partially tracing with respect to every qubit, provide us with valuable information about the
entanglement that still remains and therefore that is going to characterize every reduced subsystem.
It thus follows that the original polynomial contains all the monomials associated with all possible
maximally entangled subsystems, which can be determined by means of the Peres-Horodecki
separability criterion \cite{peres:PRL:1996,horodecki:PLA:1996} (see also \cite{cirac:PRL:1999,cirac:PRA:2000}).
This is, somehow, analogous to the criterion introduced in \cite{quinta:PRA:2018} to reduce
the polynomials under the presence of redundant monomials (simplification rule): a monomial is
irrelevant and, therefore, must be neglected if all its variables are already present in a lower-order
sub-polynomial, which corresponds to a maximally entangled state that does not include other variables.

To illustrate the above assertion, consider the case, for instance, of the quantum state
$|\Psi^{34}\rangle$ from \cite{quinta:PRA:2018} (the subscript `34' is just a classifying
label, not relevant to the present work, but introduced to establish a better connection
to notation used in the literature), which reads as
\ba
 |\Psi^{34}\rangle & = & |3^1\rangle_{abc}|1\rangle_d|0\rangle_\nu + |3^1\rangle_{abd}|0\rangle_c|1\rangle_\nu
 + |3^1\rangle_{acd}|1\rangle_b|2\rangle_\nu
 + |3^1\rangle_{bcd}|0\rangle_a|3\rangle_\nu \nonumber \\
 & & + |2^1\rangle_{ab}|11\rangle_{cd}|4\rangle_\nu
 + |2^1\rangle_{cd}|11\rangle_{ab}|5\rangle_\nu .
 \label{eq:pure}
\ea
This state describes a system formed by 4 qubits ($a$, $b$, $c$ and $d$) coupled to an auxiliary 6-level qudit~$\nu$ (specified by $|q\rangle_\nu$, with $q$ ranging from 0 to 5).
Notice that, in turn, this quantum state consists of several tripartite and bipartite entangled
sub-states of the class
\be
 |N^1\rangle_{ijk\ldots} =\frac1{\sqrt{2}} \left( |00\ldots\rangle_{ijk\ldots} + |11\ldots\rangle_{ijk\ldots} \right) ,
 \label{Nqubitent}
\ee
with $i,j,k = a,b,c,d$, i.e.,
\ba
 |2^1\rangle_{ij}  & = & \frac1{\sqrt{2}} \left( |00\rangle_{ij} + |11\rangle_{ij} \right) , \\
 |3^1\rangle_{ijk} & = & \frac1{\sqrt{2}} \left( |000\rangle_{ijk} + |111\rangle_{ijk} \right) .
 \label{eqNn}
\ea
Furthermore, there are also factorizable single and bipartite qubits (note that the difference in notation between
factorizable states and entangled ones relies on the superscript associated with the latter, which makes reference
to the corresponding class specified in \cite{quinta:PRA:2018}).
Physically, the quantum state (\ref{eq:pure}) can be interpreted as the result of coupling a four-qubit system with
a qudit in such that the latter couples differently depending on how the four qubits are arranged.
For instance, the qudit remains in its ground state if the qubits $a$, $b$ and $c$ are entangled, while the qubit $d$
remains independent; if the qubits $a$ and $b$ are entangled, and the qubits $c$ and $d$ remain
independent, then the qudit $\nu$ couples with its fourth excited state; and so forth.

\begin{figure}[!t]
 \begin{center}
  \includegraphics[width=0.50\textwidth]{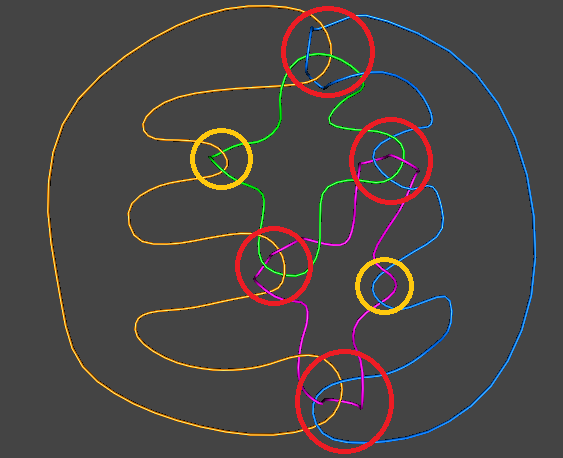}
  \caption{Topological link associated with the polynomial $\mathcal{P}=abc+abd+acd+bcd+ab+cd$, which represents the
  mixed state described by the reduced density matrix (\ref{eq:mixed}).
  The orange, green, blue and pink rings denote, respectively, qubits $a$, $b$, $c$ and $d$, with the yellow and red circles enclosing,
  respectively, bipartite and tripartite entangled structures (monomials).}
  \label{Fig1}
 \end{center}
\end{figure}

In order to determine now the amount of entanglement as well as the polynomials and topological
rings associated with (\ref{eq:pure}), we first determine the corresponding density matrix
and then we partially trace over the qudit $\nu$.
The resulting 16$\times$16 reduced density matrix reads as
\ba
 \hat{\rho}_{abcd}\left(\Psi^{34}\right) & = &
 \frac{{\rm Tr}_\nu \left[ |\Psi^{34}\rangle\langle \Psi^{34}| \right]}{\langle \Psi^{34}|\Psi^{34}\rangle}
 \nonumber \\
 & = &   |3^1\rangle_{abc} |1\rangle_d {_d\langle}1| {_{abc}\langle 3^1|}
       + |3^1\rangle_{abd} |0\rangle_c {_c\langle 0|} {_{abd}\langle 3^1|} \nonumber \\
 & & +  |3^1\rangle_{acd} |1\rangle_b {_b\langle 1|} {_{acd}\langle 3^1|}
       + |3^1\rangle_{bcd} |0\rangle_a {_a\langle 0|} {_{bcd}\langle 3^1|} \nonumber \\
 & & +   |2^1\rangle_{ab} |11\rangle_{cd} {_{cd}\langle 11|} {_{ab}\langle 2^1|}
       + |2^1\rangle_{cd} |11\rangle_{ab} {_{ab}\langle 11|} {_{cd}\langle 2^1|} .
 \label{eq:mixed}
\ea
After applying the Peres-Horodecki separability criterion for mixed states to this reduced
matrix (details on this calculation are provided in Appendix~\ref{appA}), only the subsystems
(monomials) $abcd$, $abc$, $abd$, $acd$, $bcd$, $ab$ and $cd$ are shown to be maximally
entangled.
Consequently, the associated polynomial is
\be
 \mathcal{P} = abc + abd + acd + bcd + ab + cd ,
 \label{polyn}
\ee
which contains all the above monomials, except the $abcd$ one, neglected by virtue of the
simplification rule mentioned above.
On the other hand, the topological link associated with the polynomial (\ref{polyn}) is
readily obtained by cutting and removing the ring corresponding to the qudit $\nu$, which is
equivalent to setting to zero the associated variable.
This process is analogous to chain structures (see, for instance, Fig.~11 in \cite{quinta:PRA:2018}),
each one representing a monomial with 2, 3 or 4 variables.
In the particular case of the mixed state (\ref{eq:mixed}), the corresponding link representation is
displayed in Fig.~\ref{Fig1}, where qubits $a$, $b$, $c$ and $d$ have been denoted with orange, green,
blue and pink rings, respectively, for a better visualization.
Furthermore, in order to better appreciate the monomial structure, i.e., the two- and three-variable
linking, bipartite entangled structures have been surrounded by yellow and red circles, respectively.

As it can be noticed, a topological description in terms of links is rather convenient to
specify entanglement relationships, for it provides us with a general picture of this
quantum property without making any explicit reference to a particular physical substrate
(material particles, photons or, in general, different types of degrees of freedom).
The key question now is whether the same representation is suitable to describe entanglement
transfer or exchange among different quantum systems, which is the case of quantum
teleportation.
Two drawbacks arise in this regard:
\begin{enumerate}
 \item The way back is not straightforward, i.e., it is not evident how to generate links
 associated with polynomials from the constituting monomials.
 These entails difficulties regarding the introduction of some general rules aimed at
 describing the system splitting after measurement.

 \vspace{.25cm}
 \item Although topological links can be useful to elucidate entanglement relationships in
 systems with a few qubits, the same gets far messier as the number of qubits increases.
\end{enumerate}
At this point, it is clear we need an alternative representation, which may introduce or
describe similar elements, but that, in turn, becomes more versatile, thus providing us
with a more suitable framework to deal with the notion of flow of entanglement and
information.


\section{Graphs and entanglement}
\label{sec3}

In order to circumvent the above inconveniences, let us thus consider instead a graph
representation.
In order to move from polynomials to graphs, we associate the variables of a polynomial
with (labeled) graph nodes and any entanglement relationship is accounted for by a graph
edge.
Accordingly, within this scenario, setting to zero a variable is analogous to removing the
associated node (and, of course, also the edges incident on it).
Now, this is still a rather basic representation, since it only enables a description of
polynomials consisting of two-variable monomials; higher-order monomials require an additional level of refinement, for the removal of a node and the corresponding incident edges should not
generate a disconnection among the remaining nodes.
Notice that this is also necessary, in turn, in order to provide a dynamical view of the
entanglement transfer process.

\begin{figure}[!t]
 \begin{center}
  \includegraphics[width=1.00\textwidth]{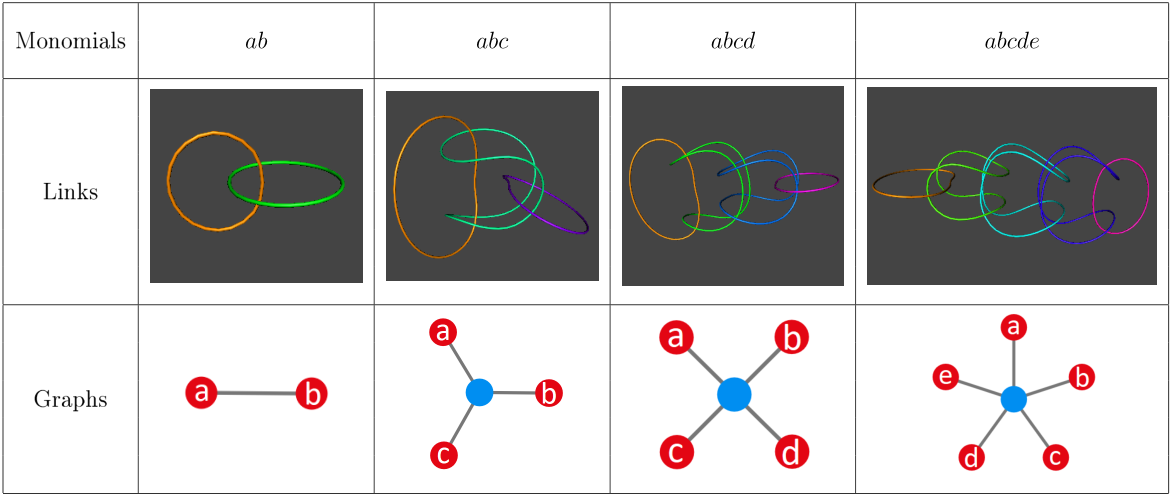}
  \caption{Monomial, link and graph representation for several $N$-qubit entangled states, with $N = 2, 3, 4$ and 5,
  which correspond, respectively, to the quantum states $|2^1\rangle$, $|3^1\rangle$, $|4^1\rangle$ and $|5^1\rangle$.
  These states are of the type exemplified by the states (\ref{eqNn}).}
  \label{Fig2}
 \end{center}
\end{figure}

That problem is solved here by defining an auxiliary element in the approach, namely an
unlabeled connection or virtual node.
Appealing to a color code for simplicity, these virtual nodes are going to be denoted here
with blue, while the usual labeled ones will be specified with red.
The constructions of graphs by means of these auxiliary virtual nodes follows some simple rules,
which also help to simplify the reconstruction of multipartite graphs from elementary ones,
just in the same that polynomials are built from elementary monomials.
These rules are:
\begin{enumerate}
 \item Auxiliary nodes are only used to join three or more labeled nodes, which represent
  entangled states for three or more parties.
  Accordingly, two (or more) auxiliary nodes cannot be directly connected by an edge.

 \vspace{.25cm}
 \item An auxiliary node is redundant if it is connected to three or more labeled nodes
  chained among themselves by successive edges.
  Redundant nodes must be removed.

 \vspace{.25cm}
 \item If a labeled node is connected to other nodes through a virtual node, the trace over
  such a node removes not only its edge with the virtual node, but also the latter.
\end{enumerate}

To illustrate the concept, an equivalence table is displayed in Fig.~\ref{Fig2}, where monomials, links
and graphs for entangled states with a different amount of parties ($N = 2$ to 5) are compared.
These states have been built following a simple recursive relation with $N$, which makes
more apparent the fact that, when one of the qubits is removed, no matter which one,
all other qubits become disentangled immediately.
That is, setting to zero a variable, cancels out completely the corresponding monomial;
cutting a ring, totally releases all other rings.
In the case of the graph representation, it can be seen that, for $N \ge 3$, the virtual
node keeps all qubits entangled.
However, as soon as one of such qubits is removed (we partially trace over its states), the
virtual node disappears and all other remaining qubits are released.
This approach thus starts pointing in a direction that seems to favor a description in terms
of generation and annihilation of entangled states, which is precisely the case in quantum
teleportation, as it will be seen below.

\begin{figure}[!t]
 \begin{center}
 \includegraphics[width=0.8\textwidth]{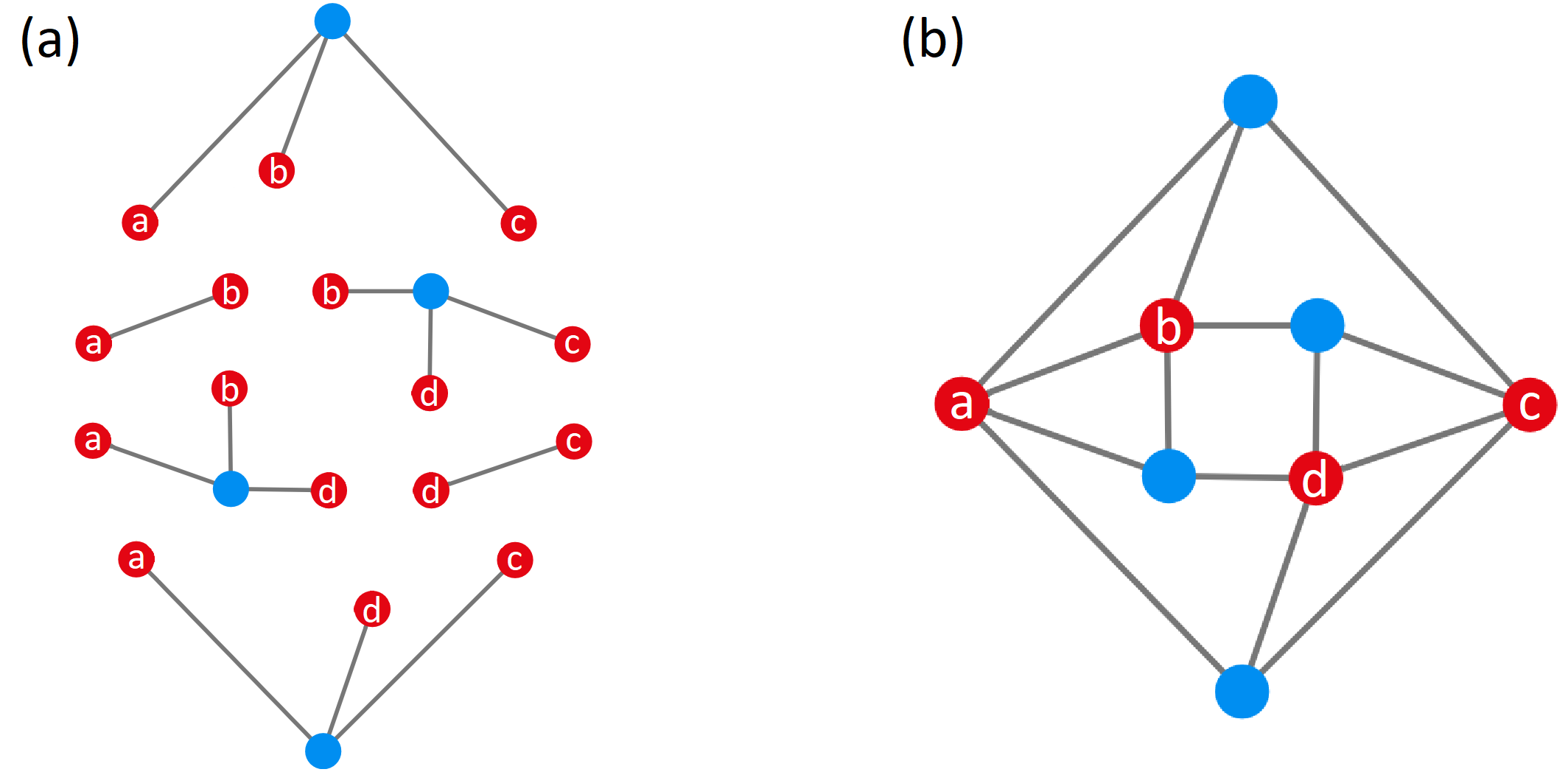}
 \caption{(a) Graph representation of the monomials constituting the polynomial (\ref{polyn}).
  (b) Graph representation of the tetrapartite entangled represented by the polynomial
  (\ref{polyn}).}
 \label{Fig3}
 \end{center}
\end{figure}

Let us now consider the construction of the graph associated with a given polynomial.
Without any loss of generality, we are going to consider the polynomial (\ref{polyn}).
The graph representation for the six constituting monomials is shown in Fig.~\ref{Fig3}(a).
These monomials have been arranged in such a way that identical elements (qubits) appear
face-to-face, which facilitates the subsequent construction of the final graph.
Notice the key role played here, within this arrangement, by the virtual nodes.
As it is seen in Fig~\ref{Fig3}(b),
these nodes do not disappear when identical elements (labeled nodes) are merged together,
because they are not redundant, according to the above rules, rather they provide us a clue
on the inverse disconnection process.
Furthermore, from this tetrapartite entangled system, the presence of the virtual nodes
allows us to easily identify entangled substructures (monomials) by just following the
edges that join them with labeled nodes (for bipartite entangled structures it is enough
to follow the edge that joins two neighbor labeled nodes).

\begin{figure}[t]
\begin{center}
\includegraphics[width=0.5\textwidth]{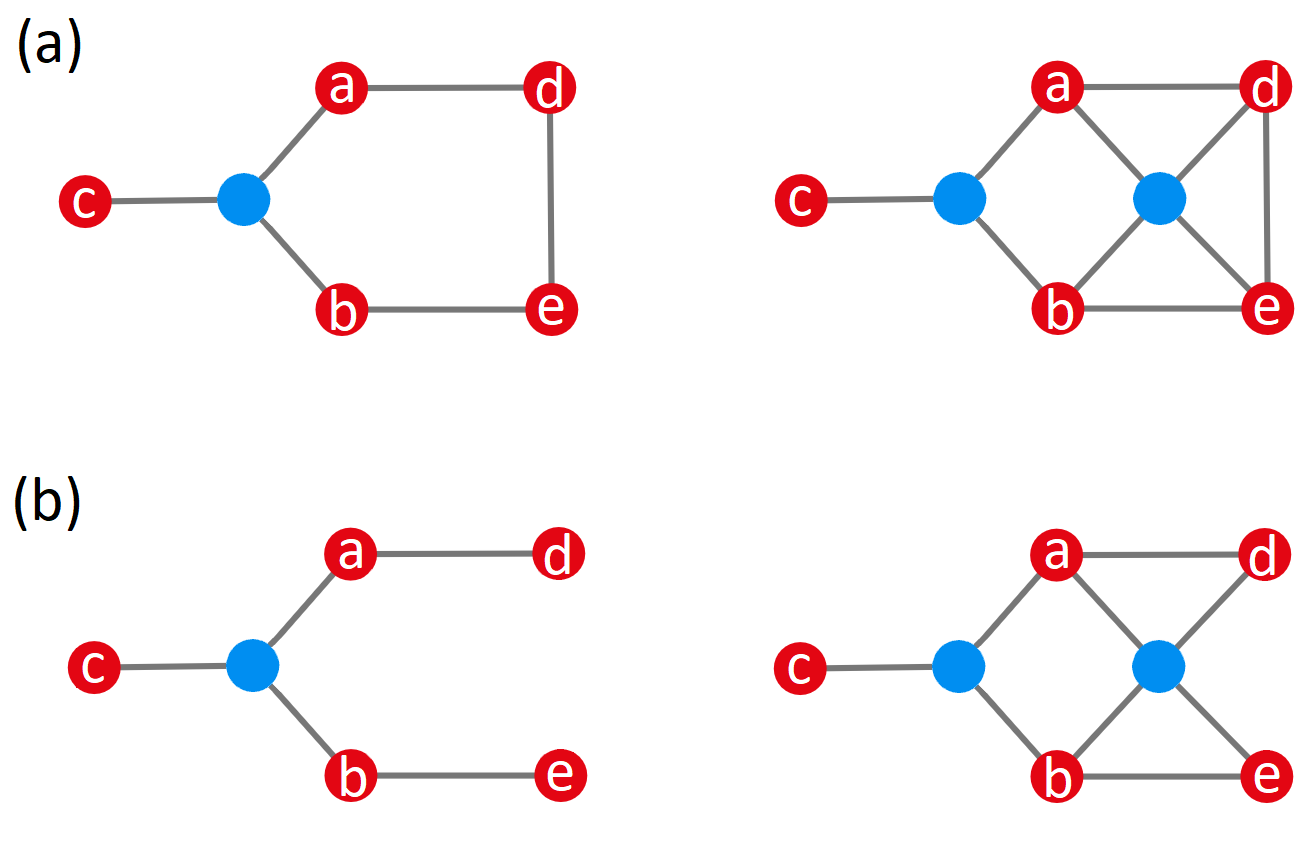}
\caption{In (a), both graphs correspond to the polynomial $\mathcal{P}=abc+ad+be+de$, and the valid graph is the one on the left. In (b), the graph on the left corresponds to the polynomial $\mathcal{P}=abc+ad+be$ and the right-hand graph corresponds to $\mathcal{P}=abde+abc+ad+be$.}
\label{Fig4}
\end{center}
\end{figure}

By virtue of the second rule above, graphs can also be simplified.
To illustrate this property, consider the pentapartite entangled state described by the
polynomial
\be
 \mathcal{P} = abc + ad + be + de .
\ee
The two graphs in Fig.~\ref{Fig4}(a) are equivalent representations of this polynomial,
although the one on the right contains a redundant auxiliary node.
Note that this node joins labeled nodes that form a chain, which can be removed without
altering the relationship among the constituting qubits, described by the monomials $ad$,
$be$ and $be$.
The other virtual node, on the contrary, is not reduntant, describing the tripartite
monimial $abc$.
Now, notice that it is enough removing a single edge, for instance, between $d$ and $e$,
to make such a virtual node not redundant, as it is illustrated by the two graph
representations displayed in Fig.~\ref{Fig4}(b).
Although they also describe pentapartite entangled systems, these graphs are not equivalent.
The graph on the left is associated with the polynomial
\be
 \mathcal{P} = abc + ad + be .
\ee
The strategic position of the second virtual node in the graph on the right, on the other
hand, additionally contains a tetrapartite entangled subsystem, namely ($abde$), thus giving
rise to the polynomial
\be
 \mathcal{P} = abde + abc + ad + be .
\ee


\section{Graphs and quantum teleportation}
\label{sec4}

In order to determine a proper graph-based description for teleportation processes
regardless of the amount of entangled qubits involved in the quantum channel, first
we are going to examine several cases, ranging from the typical two-qubit maximally
entangled pair to states that include up to 5 or 6 entangled qubits.
This will provide us with a general set of rules applicable to any case.
Nevertheless, in all cases we are analyzing scenarios where the information encoded in a
given qubit has to be transmitted to a far distant laboratory, which owns one of the parties
involved in the corresponding $N$-qubit entangled state.


\subsection{Two-qubit entangled state}
\label{sec41}

\begin{figure}[t]
\begin{center}
\includegraphics[width=0.25\textwidth]{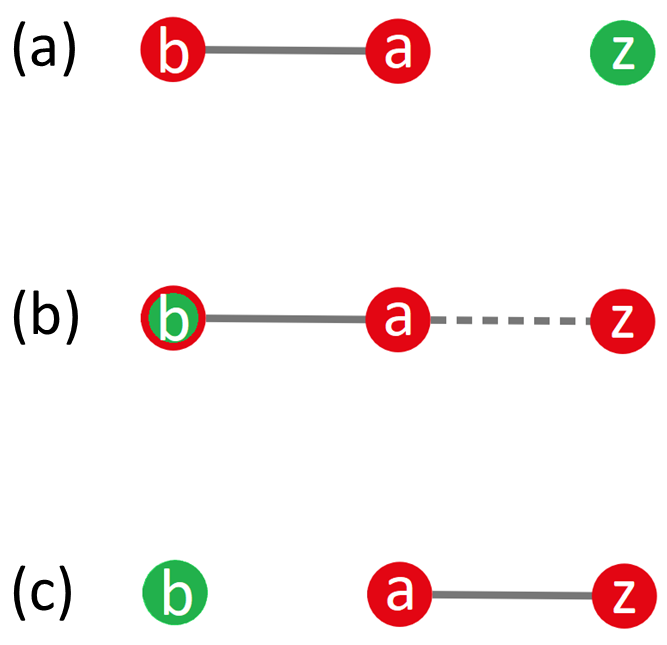}
 \caption{Three-stage graph representation of the quantum teleportation of a qubit $z$
 by means of an entangled pair $ab$:
 (a) Alice has the qubit $z$, while she also shares
 one party of an entangled pair with Bob;
 (b) the qubit $z$ interacts with Alice's
 entangled party;
 (c) the information encoded in $z$ is transferred (teleported) to
 Bob's entangled party after measurement.}
\label{Fig5}
\end{center}
\end{figure}

So far we have focused on the relational aspects involved in entangled states or subsystems.
In order to tackle quantum information exchange, as it is the case of quantum teleportation,
it is of interest first to briefly see how the process takes place.
To this end, let us reconsider the basic quantum teleportation protocol \cite{bennet:PRL:1993},
first in the usual state vector representation and then its description in the present graph
approach.
Accordingly, we have Alice ($a$) who wishes to teleport the information specified by the quantum state of a qubit ($z$) that she owns to Bob ($b$), in a far distant laboratory, by using as a
quantum channel the two parties of an entangled pair shared by them.
Let us denote the quantum state accounting for the full three-qubit system as
\be
 |\Psi\rangle_{zab} = |\phi\rangle_z|2^1\rangle_{ab} ,
\ee
which contains the maximally entangled bipartite subsystem $ab$ and the factorized single
qubit $z$.
Following the rules provided in the previous section, this state or first stage of the process
can be represented by means of the upper graph displayed in Fig.~\ref{Fig5}(a).

After Alice's entangled party interacts with her qubit, the state describing this new
two-qubit system can be recast in terms of Bell-basis vectors as
\be
 |\Psi\rangle_{zab} = |\Phi^{\left(+\right)}_{za}\rangle|\phi_b^{\left(1\right)}\rangle
  + |\Phi^{\left(-\right)}_{za}\rangle|\phi_b^{\left(2\right)}\rangle
  + |\Psi^{\left(+\right)}_{za}\rangle|\phi_b^{\left(3\right)}\rangle
  + |\Psi^{\left(-\right)}_{za}\rangle|\phi_b^{\left(4\right)}\rangle .
 \label{eq:telep2}
\ee
In this second stage, the change of basis is going to induce a series of important
modifications in the previous graph structure, as it can be noticed in the middle graph
shown in Fig.~\ref{Fig5}(b), with clear physical implications.
First, not only $a$ and $z$ are going to be joined by a virtual (dashed) edge, but $z$ itself
is also going to undergo a change in its color in order to make more apparent the new pair
$az$ defined according to the Bell basis set.
Moreover, $b$ is also redefined with green color, although surrounded by a red circle, with
the purpose to explicitly indicate its potentiality to acquire the information (state)
formerly encoded in $z$.
However, the fact that it is still physically attached (entangled) to $a$, as denoted by the
solid edge, is remarked by the red circle.

Therefore, it is only after Alice performs the measurement on her pair of entangled
particles, when Bob's qubit is forced to project over the corresponding Bell vector.
By transmission through a classical channel of two bits (operations), Alice communicates
to Bob the result of her measurement.
Now he knows the quantum state of his qubit and, by using the correct operator, he finally
gets the original state.
In terms of graphs, this physical process implies removing the ambiguity of $b$, which gets
physically separated from $a$ (no solid edge between them) and hence acquires an unambiguous green color, while $a$ and $z$ become maximally entangled (the virtual edge becomes real).
The process is described by the lower graph shown in Fig.~\ref{Fig5}(c).

\subsection{Three-qubit entangled state}
\label{sec41-3N}

\begin{figure}[t]
\begin{center}
\includegraphics[width=0.7\columnwidth]{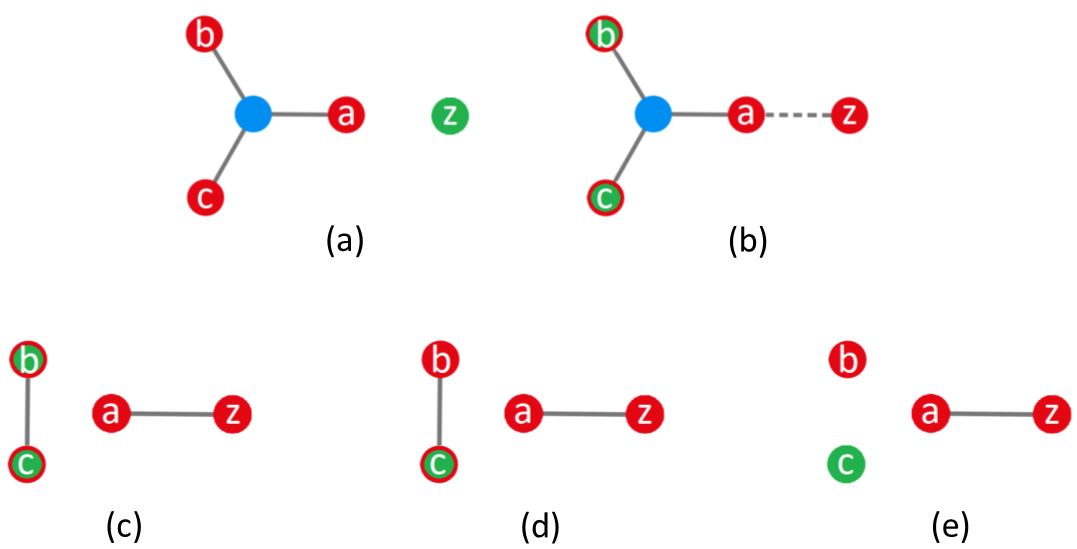}
 \caption{Graph representation of the teleportation process sharing a $|3^1\rangle$ entangled state as quantum channel (see text for a detailed account of each stage of the evolution).}
\label{Fig6}
\end{center}
\end{figure}

Using a standard two-qubit entangled state may seem sort of trivial.
Let us thus consider a case where three parties share the entangled qubits a three-qubit
$|3^1\rangle$ state, introduced above.
The teleportation process for such kind of states is formerly described in \cite{karlsson:PRA:1998}.
Now, Alice wishes to send the information encoded in $|\phi\rangle_z$ to Cliff, but taking
into account her partner Bob.
So, the three of them receive in their corresponding laboratories one of the three parties
implied in the $|3^1\rangle$ state.
As for the transfer of classical information, unlike the previous setup, this time three bits
are required, namely two from Alice to Bob and one from Bob to Cliff.
The quantum state describing the four qubits that come into play reads as
\be
 |\Psi\rangle_{zabc} = \frac{1}{\sqrt{2}} \cos (\theta/2) \Big(|0000\rangle+ |0111\rangle\Big)
  + \frac{1}{\sqrt{2}} \sin (\theta/2) e^{i\varphi} \Big(|1000\rangle+ |1111\rangle\Big) .
 \label{eq3N-1}
\ee
The graph representation for this state is provided in Fig.~\ref{Fig6}(a), where the physical separation
between the $abc$ entangled system and the $z$ qubit is evident following the description
provided in Sec.~\ref{sec41}.
Next, Alice proceeds with a change of basis to describe her qubits $z$ and $a$, which in terms of the
Bell basis vectors allows to recast (\ref{eq3N-1}) as
\ba
 |\Psi\rangle_{zabc} & = &
    \phantom{+} \frac{1}{2} |\Phi^+\rangle \left[ \cos(\theta/2) |00\rangle + \sin(\theta/2) e^{i\varphi} |11\rangle \right]
  + \frac{1}{2} |\Phi^-\rangle \left[ \cos(\theta/2) |00\rangle - \sin(\theta/2) e^{i\varphi} |11\rangle \right] \nonumber \\
 & & + \frac{1}{2} |\Psi^+\rangle \left[ \sin(\theta/2) e^{i\varphi} |00\rangle + \cos(\theta/2) |11\rangle \right]
  + |\Psi^-\rangle \left[- \sin(\theta/2) e^{i\varphi} |00\rangle + \cos(\theta/2)|11\rangle \right] . \nonumber \\ & &
 \label{eq3N-2}
\ea
The graph representation for this rewritten state is given by Fig.~\ref{Fig6}(b), where the
link between qubits $a$ and $z$ is again denoted by the dashed line, while the potentiality
to receive the information encoded in $z$ is denoted by the green labeled nodes encircled
by the red (qubits $b$ and $c$).

Once the tetrapartite state (\ref{eq3N-1}) is expressed in terms of the Bell basis set,
it is clear that a measurement by Alice on $a$ and $z$ will render any of the four basis
vectors with the same probability, releasing the other two qubits, $b$ and $c$, from their
entanglement with $a$.
These vectors become entangled between themselves, being described by a state $|2^1\rangle$.
This is precisely the situation described by the graph displayed in Fig.~\ref{Fig6}(c),
where the virtual node has disappeared (it is not necessary because we have two separate
bipartite entangled states), and the qubits $b$ and $c$ still undefined, as it is denoted
by the green-in-red nodes.
Remember that, in the case discussed in the previous section, after measurement, the released
qubit had already acquired the information initially encoded by $z$ (thus becoming green).
Such undefined state stresses the potentiality for any of these two qubits to acquire the
state of $z$.

The next step is rather subtle.
Without any loss of generality, let us assume that, after measurement, Alice obtains that
the system $za$ is in the state $|\Phi^+\rangle$.
This means that $bc$ are in the entangled state
\be
 |\Psi\rangle_{bc} = \cos (\theta/2) |00\rangle + e^{i\varphi} \sin (\theta/2)|11\rangle .
 \label{eq3N-3}
\ee
So, the information encoded in $z$ is shared by $b$ and $c$, but neither Bob nor Cliff have
it yet.
As it can be noticed, a measurement on $b$ would not solve the problem, for it would imply
that Cliff's qubit, $c$, would be either in the state $|0\rangle$ or in $|1\rangle$, with
the corresponding probabilities, but we would not recover the initial state $|\phi\rangle$.
Now, in the same way that in order to reach (\ref{eq3N-3}) Alice had to consider a change
of basis, the same can be done by Bob, not to a Bell basis, but to a basis set rotated by
an amount with respect to the $\{|0\rangle,|1\rangle\}$.
Let us denote this new basis set for Bob $\{|+\rangle,|-\rangle\}$, which is related to the
old one by means of the transformation
\ba
 |1\rangle_b & = & \cos \omega |+\rangle_b + \sin \omega |-\rangle_b , \nonumber \\
 |0\rangle_b & = & -\sin \omega |+\rangle_b + \cos \omega |-\rangle_b , \nonumber
\ea
where $\omega$ is the rotation angle,  which can be achieved by means of
a Stern-Gerlach magnets for entangled charged particles or polarizers for photons
(this angle could be taken equal to 45$^\circ$, for instance).
Accordingly, the state for $bc$ can be written as
\ba
 \label{eq:ecpsibc}
 |\Psi\rangle_{bc} & = & \left[ - \sin \omega \cos (\theta/2) |0\rangle_c + \cos \omega
 \sin (\theta/2) e^{i\varphi}  |1\rangle_c \right] |+\rangle_b \nonumber \\
 & & + \left[ \cos \omega \cos (\theta/2) |0\rangle_c + \sin \omega \sin (\theta/2) e^{i\varphi} |1\rangle_c \right] |-\rangle_b ,
\ea
which is represented by the graph shown in Fig.~\ref{Fig6}d.
As it can be noticed in this figure, the ambiguity has disappeared from qubit $b$; all
potentiality relies on qubit $c$.

The last step thus consist of Bob performing a measurement of his qubit in the new basis,
and trasmitting classically the result to Cliff, so that he can then proceed to specify
the state of his qubit.
Actually, depending on such a result, Cliff should apply one of the following operators.
If the outcome is $|+\rangle_b$, then he should operate his final state by
\be
 \hat{O}_+ = \left( \begin{array}{cc}
  1/\cos \omega & 0 \\ 0 & -1/\sin \omega
 \end{array} \right) ,
\ee
while for $|-\rangle_b$ it should be used
\be
 \hat{O}_- = \left( \begin{array}{cc}
  1/\sin \omega & 0 \\ 0 & 1/\cos \omega
 \end{array} \right) ,
\ee
this being in analogy to the operators needed to recover the transmitted state in the
standard two-qubit protocol.
With this last operator, qubit $c$ is disambiguated and acquires the information formerly
encoded in qubit $z$, as denoted by the graph represented in Fig.~\ref{Fig6}(e).


\subsection{Four-qubit entangled state}
\label{sec41-4N}

\begin{figure}[t]
\begin{center}
\includegraphics[width=0.7\columnwidth]{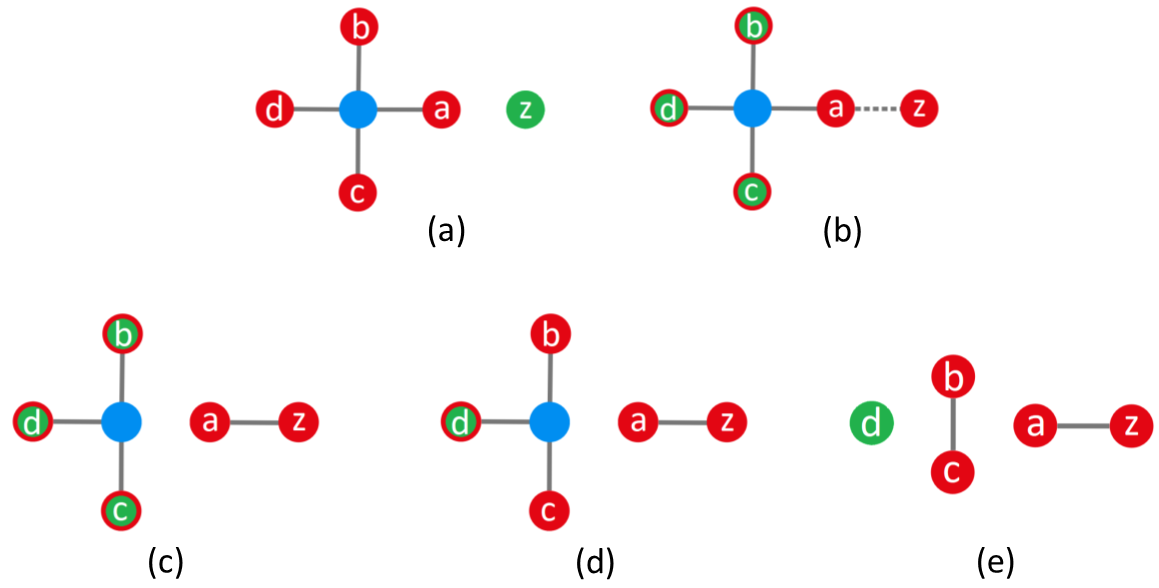}
 \caption{Graph representation of the teleportation process sharing a $|4^1\rangle$ entangled state as quantum channel (see text for a detailed account of each stage of the evolution).}
\label{Fig7}
\end{center}
\end{figure}

The same procedure described in Sec.~\ref{sec41-4N} can be now readily extended to
entangled states of the type $|N^1\rangle$ if up to $N$ parties are involved in the sharing
of information.
To illustrate it, let us consider $N = 4$, where now the information is going to be conveyed
from Alice to Dave, the fourth party, with Bob and Cliff playing the passive role of conveyors.
In this case, the total state is described by the combination $|\phi\rangle_z$ and $|4^1\rangle$,
which reads as
\be
 |\Psi\rangle_{zabcd} =
 \frac{1}{\sqrt{2}} \cos (\theta/2) \Big(|00000\rangle + |01111\rangle \Big) +
 \frac{1}{\sqrt{2}} \sin (\theta/2) e^{i\varphi} \Big(|10000\rangle + |11111\rangle \Big) ,
 \label{eq4N-1}
\ee
which is represented by the graph plotted in Fig.~\ref{Fig7}(a).
After Alice performs the change of basis for qubits $z$ and $a$, this state is
rearranged in terms of the Bell basis vectors as
\ba
 |\Psi\rangle_{zabcd} & = &
 \frac{1}{2} |\Phi^+\rangle \left[ \cos (\theta/2) |000\rangle + \sin (\theta/2) e^{i\varphi} |111\rangle \right] \nonumber \\
 & & + \frac{1}{2} |\Phi^-\rangle \left[ \cos (\theta/2) |000\rangle - e^{i\varphi} \sin (\theta/2) |111\rangle \right] \nonumber \\
 & & + \frac{1}{2} |\Psi^+\rangle \left[ \sin (\theta/2) e^{i\varphi} |000\rangle + \cos (\theta/2) |111\rangle \right] \nonumber \\
 & & + \frac{1}{2} |\Psi^-\rangle \left[ - \sin (\theta/2) e^{i\varphi} |000\rangle + \cos (\theta/2) |111\rangle \right] ,
\label{eq4N-2}
\ea
which is denoted by the graph shown in Fig.~\ref{Fig7}(b).
As before, as soon as the relation between qubits $z$ and $a$ is set by means of the virtual
node (physically, through the above basis change), all other nodes acquire the potentiality
to receive the information encoded in $z$, thus changing their color to the undefined
green-in-red one.

When Alice performs the measurement of a Bell state, the $za$ bipartite system becomes
physically disentangled from the rest, as it is shown in Fig.~\ref{Fig7}(c).
If, for instance, Alice measures $|\Phi^+\rangle$, the state describing the subsequent
entangled $bcd$ tripartite system is
\be
 |\Psi\rangle_{bcd} = \cos (\theta/2) |000\rangle + \sin (\theta/2) e^{i\varphi} |111\rangle .
 \label{eq4N-3}
\ee
Notice now that we can apply the same rotation recipe used in the previous section, though
sequentially, that is, first Bob changes the basis of his qubit and then passes the information
onto Cliff, who, in turn, proceeding the same way will allow Dave to obtain in his qubit
the information encoded in $z$.
This is an option, of course.
However, if it happens that Bob and Cliff are in the same laboratory, there is another option
that, in turn, allows to preserve a maximally entangled state, which consists in recasting
$bc$ in terms of a Bell basis set, so that (\ref{eq4N-3}) can be written as
\be
 |\Psi\rangle_{bcd} = \frac{1}{\sqrt{2}} |\Phi^+\rangle \left[ \cos (\theta/2) |0\rangle + \sin (\theta/2) e^{i\varphi}  |1\rangle \right]
 + \frac{1}{\sqrt{2}} |\Phi^-\rangle \left[ \cos (\theta/2) |0\rangle - \sin (\theta/2) e^{i\varphi} |1\rangle \right] ,
\ee
represented by the graph in Fig.~\ref{Fig7}(d).
However, noticed that, unlike Alice's former step, now only two Bell vectors are involved,
which simplifies the measurement process.

So, from this point on, the procedure is analogous to the one seen in the previous section,
for it is only necessary a measurement by Bob and Cliff, and the transmission of the outcome
(two bits) by a classical channel, which will be enough for Dave to obtain the replica of
$z$ in his qubit, as shown in Fig.~\ref{Fig7}(e).
Specifically, if the outcome is $|\Phi^+\rangle$, it means $d$ has automatically acquired
the former state of $z$; if the outcome is $|\Phi^-\rangle$, the projective operator
\be
 \begin{pmatrix} 1 & 0 \\ 0 & -1 \end{pmatrix}
\ee
must be used before.


\subsection{General $N$-qubit entangled states}
\label{sec42}

The cases analyzed in Secs.~\ref{sec41-3N} and \ref{sec41-4N} provide us with some
insight on a general procedure to dynamically describe quantum teleportation processes
involving $N$-qubit entangled states as a quantum resource.
Two sequential routes have been mentioned.
From the analysis of teleportation involving $|3^1\rangle$ states, we found that after
the measurement carried out by Alice, the transfer of information is achieved by simple
rotations.
This procedure is independent on how many qubits are involved in the quantum channel,
which makes possible effective transfers among different laboratories, since each party
from the entangled system is allocated in one of them.
The other procedure, coming out from the analysis including $|4^1\rangle$ states, consisted
in transferring the information by producing two-party rotations in a reduced Bell basis.
This procedure is optimal if qubits can be allocated two by two within the same laboratory,
which has the advantage that, after transmission, those two qubits are maximally entangled
and can be used eventually as a quantum resource for further operations.
Of course, there can also be hybrid options, where some laboratories receives only one party,
while others receive two of them.

In all of this, though, we can find or set a kind of general protocol in terms of a graph
representation, which allows us to better understand the different steps of the process,
thus providing us with an insightful dynamical picture.
This protocol involves a series of common steps, even if there are some variations.
Thus, without any loss of generality, let us consider a simple situation, where we can
pair as many as possible two-qubit entangled parties.
All intermediate steps are going to be the same regardless of $N$, but not the last one,
which will depend on whether $N$ is odd or even, but this is something we shall specify
in due time.
So, given a quantum channel ascribed to an $N$-party entangled system, with its quantum
state denoted by $|N^1\rangle$, as given by (\ref{Nqubitent}), the graph representation
of quantum teleporting the state $|\phi\rangle_z$ to any of the $N$ parties includes a
series of steps, each one accounting for a specific operation.
In order to illustrate these steps, we are going to consider the processes illustrated in
Figs.~\ref{Fig8} and \ref{Fig9}, where the corresponding teleportation processes involved
$|5^1\rangle$ and $|6^1\rangle$ states, respectively.
Accordingly, we have the following steps:
\begin{enumerate}
 \item {\bf Initial preparation.}
  The graph for the $|N^1\rangle$ state is denoted by $N$ red labeled nodes ($a, b, c, d, e, f, g, \ldots$), all of them joined by solid edges with a virtual (blue) node, which
  represents the entangling correlation.
  In turn, the state $|\phi\rangle_z$ is represented by a green isolated (labeled) node.
  This is what we observe in part (a) of Figs.~\ref{Fig8} and \ref{Fig9}.

\begin{figure}[t]
\begin{center}
\includegraphics[width=0.9\textwidth]{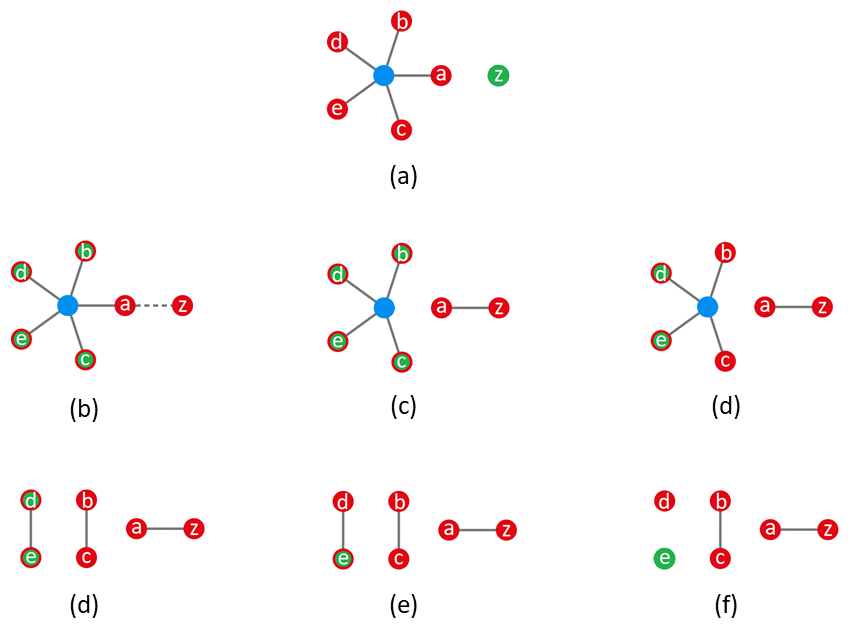}
 \caption{Graph representation of the teleportation process sharing a $|5^1\rangle$ entangled state as quantum channel (see text for a detailed account of each stage of the evolution).}
\label{Fig8}
\end{center}
\end{figure}

\vspace{.25cm}
 \item {\bf Initial basis change.}
  Alice prepares her two parties, the one from the $|N^1\rangle$ state ($a$) and $|\phi\rangle_z$,
  in a way that, when a Bell measurement is performed on this pair, one of the four Bell
  states is obtained.
  This introduces in our graph representation some interesting changes.
  First, such redefinition gives rise to the appearance of a virtual edge between $z$ and $a$,
  and the change of color of the former.
  This indicates that these two parties are going to become entangled, even though $a$ still
  remains attached (entangled) to the remaining $N-1$ parties by virtue of the virtual node.
  Second, because those other $N-1$ parties are susceptible to receive the state encoded in
  $z$, they are represented with green color, remembering the former representation for $z$.
  However, since they still remain entangled among themselves, there is a red circle around them.
  This stage is shown in Figs.~\ref{Fig8}(b) and \ref{Fig9}(b).

\vspace{.25cm}
 \item {\bf Alice measurement.} As a result of the measurement performed by Alice, the $za$
  system disentangles from the rest.
  Accordingly, the virtual (dashed) mode joining $z$ and $a$ becomes real (solid).
  The remaining $N-1$ parties are entangled among themselves, but not maximally due to the
  information transferred to them.
  The corresponding $|(N-1)^1\rangle$ state is of the same kind as (\ref{Nqubitent}), which
  readily suggest a description of every two constituting qubits in terms of the reduced
  Bell basis set $\{\Phi^+, \Phi^-\}$.
  This is the situation described by the graphs in Figs.~\ref{Fig8}(c) and \ref{Fig9}(c).

\vspace{.25cm}
 \item {\bf Two-by-two bipartite disentangling.}
  Next, two qubits from the $|(N-1)^1\rangle$ state are grouped and their state is separately
  described in terms of any of the two vectors of the reduced Bell basis.
  This implies a change of their color to red, because they are assumed to have lost their
  potentiality to received the information from $z$.
  This situation is represented in Fig.~\ref{Fig8}(d), but also in Figs.~\ref{Fig9}(d) and (f).
  In order to complete the disconnection of these pairs of qubits from the rest, a
  measurement is performed on the reduced Bell basis; the information is conveyed (by a
  classical channel) to the receiver of the state encoded in $z$ (in the examples of
  Figs.~\ref{Fig8} and \ref{Fig9}, this role corresponds to either Emily or Fred, respectively).
  Such a measurement generates bipartite maximally entangled states (which can be later
  on used, in turn, as a quantum resource), as it is seen in Figs.~\ref{Fig8}(e), and
  \ref{Fig9}(e) and (g).

\vspace{.25cm}
 \item {\bf Single-qbit measurement (only for odd $N$).}
  As it can be noticed in Fig.~\ref{Fig9}(e), when $|N^1\rangle$ involves an odd number of
  qubits, after the two-qubit disentangling operations, there are two remaining qubits, one
  of them being the one of interest (Emily's one, in this case).
  This requires, as it was seen in Sec.~\ref{sec41-3N}, a simple basis rotation and
  measurement within such a basis for the residual qubit, represented by the graph shown
  in Fig.~\ref{Fig9}(f).

\begin{figure}[t]
\begin{center}
\includegraphics[width=0.9\textwidth]{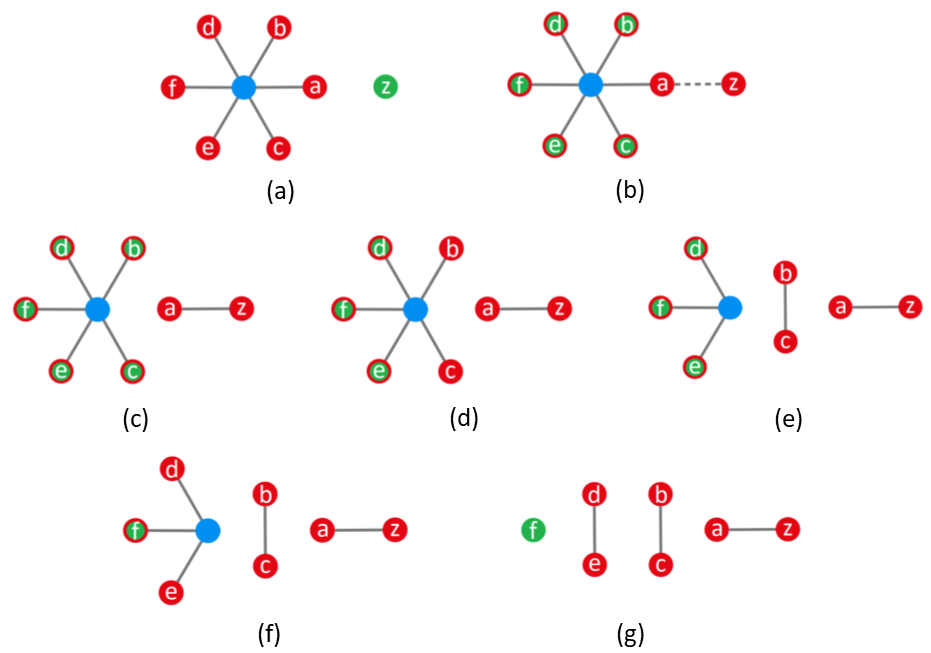}
 \caption{Graph representation of the teleportation process sharing a $|6^1\rangle$ entangled state as quantum channel (see text for a detailed account of each stage of the evolution).}
\label{Fig9}
\end{center}
\end{figure}

\vspace{.25cm}
 \item {\bf Completion of the transfer (teleportation) process.}
  The outcomes obtained after measuring in the last reduced Bell basis (for even $N$) or
  in the rotated basis (for odd $N$) will be transmitted through a classical channel to
  either Fred or Emily, who proceeding with the corresponding quantum gate operators on
  their qubit will finally obtain the state encoded in $z$, as seen in Figs.~\ref{Fig8}(g)
  and \ref{Fig9}(g).
\end{enumerate}
As it can be noticed, apart from conveying the state encoded in Alice's qubit $z$ to
the far distant qubits of Emily ($e$) or Fred ($f$), we end up with $(N-1)/2$ or $N/2$,
respectively, bipartite maximally entangled states of the type $|\Phi^\pm\rangle$, which can be
later on used as quantum resources for other purposes.


\section{Final remarks}
\label{sec5}

In spite of power of the standard algebraic approach of quantum mechanics in the field of 
quantum information transfer processes, it is equally important to have at hand simpler and 
visual approaches that emphasize the physics involved in such processes from a quick glance, 
particularly in those cases where many different relations can be established among a rather 
large number of intervening parties.
In other words, models and descriptions that emphasize the relational aspects of such complex 
systems are at the same footing, in this regard, as tough algebraic representations, which are 
more susceptible to hide the physics that otherwise we wish to readily see.
Just to establish a direct and clear analogy, think of a roadmap that allows us to quickly
see the different roads that join some places with others without involving more quantitative
measures (distances, heights, weather conditions, etc.), which is what matters at a first
glance (other quantitative details are also important, of course, but on the next level).

In that sense, topological and pictorial representations of quantum entanglement and related
processes have received much attention, because they emphasize such relational aspects over
more quantitative algebraic ones.
Now, it happens that they usually make
emphasis on a static description, suitable for the classification and description
of entangled states, or the performance and visualization of certain quantum logic
operations.
However, quantum processes, such as teleportation, involve a sort of dynamics where
entanglement is swapped from one system or subsystem into another.
In order to provide a symbolic representation for these dynamics, here we have
introduced a graph representation of the phenomenon.
With such a purpose, some additional elements, namely virtual nodes and edges, have had
to be defined in order to capture the physical essence involved in such swapping process
and translate it into proper and well-defined graph connections and disconnections.

By means of this approach, we have shown that it is possible to reproduce the flow-type
evolution involved in quantum teleportation processes, but also to recover at any stage
the usual state vector description because of the one-to-one correspondence between
both scenarios, the symbolic and the algebraic.
To prove this versatility, we have analyzed and discussed the case of state teleportation
using bipartite, tripartite and tetrapartite maximally entangled states as quantum channels,
and where the corresponding qubits could be accommodated within the same laboratory (by
pairs) or in different ones.
This has allowed us to summarize the actions that should be taken in the case of using
quantum channels made of any number of qubits with the purpose to share the information
available in a given laboratory with any other laboratory forming part of a quantum
interconnected network.


\vspace{12pt}


\noindent
{\bf Funding:} This research has been funded by the Spanish Agencia Estatal de Investigaci\'on (AEI)
and the European Regional Development Fund (ERDF) grant number FIS2016-76110-P.


\conflictsofinterest{Authors declare no conflict of interest.}


\appendixtitles{yes} 
\appendix

\section{Entanglement analysis of the reduced density matrix (\ref{eq:mixed})}
\label{appA}

The explicit form for the density matrix (\ref{eq:mixed}), $\hat{\rho}\left(\Psi^{34}\right)$, is given by
\be
 \hat{\rho}_{abcd} = \left( \begin{array}{cccccccc|cccccccc}
 2 & 0 & 0 & 0 & 0 & 0 & 0 & 1 & 0 & 0 & 0 & 0 & 0 & 1 & 0 & 0\\
 0 & 1 & 0 & 0 & 0 & 0 & 0 & 0 & 0 & 0 & 0 & 0 & 0 & 0 & 0 & 1\\
 0 & 0 & 0 & 0 & 0 & 0 & 0 & 0 & 0 & 0 & 0 & 0 & 0 & 0 & 0 & 0\\
 0 & 0 & 0 & 1 & 0 & 0 & 0 & 0 & 0 & 0 & 0 & 0 & 0 & 0 & 0 & 1\\
 0 & 0 & 0 & 0 & 1 & 0 & 0 & 0 & 0 & 0 & 0 & 0 & 0 & 0 & 0 & 1\\
 0 & 0 & 0 & 0 & 0 & 0 & 0 & 0 & 0 & 0 & 0 & 0 & 0 & 0 & 0 & 0\\
 0 & 0 & 0 & 0 & 0 & 0 & 0 & 0 & 0 & 0 & 0 & 0 & 0 & 0 & 0 & 0\\
 1 & 0 & 0 & 0 & 0 & 0 & 0 & 1 & 0 & 0 & 0 & 0 & 0 & 0 & 0 & 0\\ \hline
 0 & 0 & 0 & 0 & 0 & 0 & 0 & 0 & 0 & 0 & 0 & 0 & 0 & 0 & 0 & 0\\
 0 & 0 & 0 & 0 & 0 & 0 & 0 & 0 & 0 & 0 & 0 & 0 & 0 & 0 & 0 & 0\\
 0 & 0 & 0 & 0 & 0 & 0 & 0 & 0 & 0 & 0 & 0 & 0 & 0 & 0 & 0 & 0\\
 0 & 0 & 0 & 0 & 0 & 0 & 0 & 0 & 0 & 0 & 0 & 0 & 0 & 0 & 0 & 0\\
 0 & 0 & 0 & 0 & 0 & 0 & 0 & 0 & 0 & 0 & 0 & 0 & 1 & 0 & 0 & 1\\
 1 & 0 & 0 & 0 & 0 & 0 & 0 & 0 & 0 & 0 & 0 & 0 & 0 & 1 & 0 & 0\\
 0 & 0 & 0 & 0 & 0 & 0 & 0 & 0 & 0 & 0 & 0 & 0 & 0 & 0 & 0 & 0\\
 0 & 1 & 0 & 1 & 1 & 0 & 0 & 0 & 0 & 0 & 0 & 0 & 1 & 0 & 0 & 4\\
 \end{array} \right) ,
 \label{matrix1}
\ee
where we have removed the normalizing prefactor $1/12$ without loss of generality,
since here we are only interested in the sign of the eigenvalues.
So, after computing all the associated partial transpose matrices in order to determine the amount
of entanglement, we obtain the following sets of eigenvalues for each one:
\be
 \hat{\rho}_{abcd}^{T_a},\ \hat{\rho}_{abcd}^{T_c}:\ \left\{\begin{array}{l}
  \lambda_1 = \phantom{-}0 \ (\times 5) \\
  \lambda_2 = -1 \\
  \lambda_3 = \phantom{-}1\ (\times 5) \\
  \lambda_4 = \phantom{-}3 \\
  \lambda_5 \approx -1.306 \\
  \lambda_6 \approx \phantom{-}0.473 \\
  \lambda_7 \approx \phantom{-}1.492 \\
  \lambda_8 \approx \phantom{-}4.341
 \end{array}\right. , \qquad
 \hat{\rho}_{abcd}^{T_b},\ \hat{\rho}_{abcd}^{T_d}:\ \left\{\begin{array}{l}
  \lambda_1\phantom{0} = \phantom{-}0\ (\times 5) \\
  \lambda_2\phantom{0} = \phantom{-}1\ (\times 2) \\
  \lambda_3\phantom{0} = \phantom{-}2 \\
  \lambda_4\phantom{0} \approx-1.247 \\
  \lambda_5\phantom{0} \approx-0.887 \\
  \lambda_6\phantom{0} \approx \phantom{-}0.445 \\
  \lambda_7\phantom{0} \approx \phantom{-}0.485 \\
  \lambda_8\phantom{0} \approx \phantom{-}0.733 \\
  \lambda_9\phantom{0} \approx \phantom{-}1.802 \\
  \lambda_{10}\approx \phantom{-}2.069 \\
  \lambda_{11}\approx \phantom{-}4.600
 \end{array}\right. ,
%
%
 \nonumber
\ee
\be
 \hat{\rho}_{abcd}^{T_{ab}}:\ \left\{\begin{array}{l}
  \lambda_1 = \phantom{-}0\ (\times 8) \\
  \lambda_2 = \phantom{-}1\ (\times 2) \\
  \lambda_3 = \phantom{-}2\ (\times 2) \\
  \lambda_4 \approx -0.414 \\
  \lambda_5 \approx \phantom{-}2.414 \\
  \lambda_6 \approx -0.646 \\
  \lambda_7 \approx \phantom{-}4.646
 \end{array}\right. , \
%
 \hat{\rho}_{abcd}^{T_{ac}}:\ \left\{\begin{array}{l}
  \lambda_1\phantom{0} = \phantom{-}0\ (\times 4) \\
  \lambda_2\phantom{0} = \phantom{-}1\ (\times 4) \\
  \lambda_3\phantom{0} = \phantom{-}2 \\
  \lambda_4\phantom{0} = \phantom{-}4 \\
  \lambda_5\phantom{0} \approx -0.618 \\
  \lambda_6\phantom{0} \approx \phantom{-}1.618 \\
  \lambda_7\phantom{0} \approx -2.090 \\
  \lambda_8\phantom{0} \approx -0.356 \\
  \lambda_9\phantom{0} \approx \phantom{-}1.190 \\
  \lambda_{10} \approx \phantom{-}2.256
 \end{array}\right. , \
%
 \hat{\rho}_{abcd}^{T_{ad}}:\ \left\{\begin{array}{l}
  \lambda_1 = \phantom{-}0\ (\times 4) \\
  \lambda_2 = \phantom{-}1\ (\times 2) \\
  \lambda_3 = \phantom{-}2 \\
  \lambda_4 = \phantom{-}4 \\
  \lambda_5 \approx -1.194\ (\times 2) \\
  \lambda_6 \approx -0.295\ (\times 2) \\
  \lambda_7 \approx \phantom{-}1.295\ (\times 2) \\
  \lambda_8 \approx \phantom{-}2.194\ (\times 2)
 \end{array}\right. .
 \nonumber
\ee
As it is seen, all cases contain negative eigenvalues, which ensures the existence of
entanglement, although now we have to determine whether it is tripartite, bipartite or both,
and, if so, which parties are involved.


To proceed that way, first we have to compute the reduced matrices with respect to one
of the parties.
Thus, partially tracing over $d$ renders
\be
 \hat{\tilde{\rho}}_{abc} = \left(\begin{array}{cccc|cccc}
 3 & 0 & 0 & 0 & 0 & 0 & 0 & 1 \\
 0 & 1 & 0 & 0 & 0 & 0 & 0 & 1 \\
 0 & 0 & 1 & 0 & 0 & 0 & 0 & 0 \\
 0 & 0 & 0 & 1 & 0 & 0 & 0 & 0 \\ \hline
 0 & 0 & 0 & 0 & 0 & 0 & 0 & 0 \\
 0 & 0 & 0 & 0 & 0 & 0 & 0 & 0 \\
 0 & 0 & 0 & 0 & 0 & 0 & 2 & 0 \\
 1 & 1 & 0 & 0 & 0 & 0 & 0 & 4
 \end{array} \right)
 \label{matrix2}
\ee
with eigenvalues:
\be
 \hat{\tilde{\rho}}_{abc}^{T_a} ,\ \hat{\tilde{\rho}}_{abc}^{T_b}:\ \left\{\begin{array}{l}
  \lambda_1 = \phantom{-}0 \\
  \lambda_2 = -1 \\
  \lambda_3 = \phantom{-}1\ (\times 2) \\
  \lambda_4 = \phantom{-}2\ (\times 2) \\
  \lambda_5 = \phantom{-}3 \\
  \lambda_6 = \phantom{-}4
 \end{array}\right. , \qquad
%
%
 \hat{\tilde{\rho}}_{abc}^{T_c}:\ \left\{\begin{array}{l}
  \lambda_1 = 0\ (\times 2) \\
  \lambda_2 = 1\ (\times 2) \\
  \lambda_3 = 3 \\
  \lambda_4 \approx 0.186 \\
  \lambda_5 \approx 2.471 \\
  \lambda_6 \approx 4.343
 \end{array}\right. .
 \nonumber
\ee
Now, not all cases present negative eigenvalues, which means that tripartite
entanglement is not warranted for the $abc$ system.
Accordingly, we compute the eigenvalues of (\ref{matrix2}):
\be
 \left\{\begin{array}{lcl}
 \lambda_1 = 0\ (\times 2) & \Rightarrow & \left\{
  \begin{array}{l}
  |v_{1,1}\rangle=|100\rangle \\
  |v_{1,2}\rangle=|101\rangle
  \end{array}\right. \\
 \lambda_2 = 1\ (\times 2) & \Rightarrow & \left\{
  \begin{array}{l}
  |v_{2,1}\rangle=|010\rangle \\
  |v_{2,2}\rangle=|011\rangle
  \end{array}\right. \\
 \lambda_3 = 2 & \Rightarrow &
 |v_3\rangle=|110\rangle \\
 \lambda_4 \approx 0.657 & \Rightarrow & |v_4\rangle = -0.427|000\rangle-2.916|001\rangle+|111\rangle \\
 \lambda_5 \approx 2.529 & \Rightarrow & |v_5\rangle = -2.125|000\rangle+0.654|001\rangle+|111\rangle \\
 \lambda_6 \approx 4.814 & \Rightarrow & |v_6\rangle = \phantom{-}0.551|000\rangle-0.262|001\rangle+|111\rangle
 \end{array}\right. .
 \nonumber
\ee
Here, we observe that there are at least three non-factorizable eigenvectors,
which ensures the existence of tripartite entanglement for the $abc$ system.

If now we trace (\ref{matrix1}) over the qubit $c$, we obtain the reduced matrix
\be
 \hat{\tilde{\rho}}_{abd} = \left(\begin{array}{cccc|cccc}
 2 & 0 & 0 & 0 & 0 & 0 & 0 & 1 \\
 0 & 2 & 0 & 0 & 0 & 0 & 0 & 1 \\
 0 & 0 & 1 & 0 & 0 & 0 & 0 & 0 \\
 0 & 0 & 0 & 1 & 0 & 0 & 0 & 0 \\ \hline
 0 & 0 & 0 & 0 & 0 & 0 & 0 & 0 \\
 0 & 0 & 0 & 0 & 0 & 0 & 0 & 0 \\
 0 & 0 & 0 & 0 & 0 & 0 & 1 & 0 \\
 1 & 1 & 0 & 0 & 0 & 0 & 0 & 5
 \end{array} \right) ,
 \label{matrix3}
\ee
with the eigenvalues of its associated partial transpose matrices being
\be
 \hat{\tilde{\rho}}_{abd}^{T_a} ,\ \hat{\tilde{\rho}}_{abd}^{T_b}:\ \left\{\begin{array}{l}
 \lambda_1 = \phantom{-}0 \\
 \lambda_2 = -1 \\
 \lambda_3 = \phantom{-}1\ (\times 2) \\
 \lambda_4 = \phantom{-}2\ (\times 3) \\
 \lambda_5 = \phantom{-}5
 \end{array}\right. , \qquad
%
%
 \hat{\tilde{\rho}}_{abd}^{T_d}:\ \left\{\begin{array}{l}
 \lambda_1 = 0\ (\times 2) \\
 \lambda_2 = 1\ (\times 2) \\
 \lambda_3 = 2 \\
 \lambda_4 \approx 0.319 \\
 \lambda_5 \approx 2.358 \\
 \lambda_6 \approx 5.323
 \end{array}\right. .
 \nonumber
\ee
Again, one partial transpose has all its eigenvalues positive, so we need to compute
the eigenvectors,
\be
 \left\{\begin{array}{lcl}
 \lambda_1 = 0\ (\times 2) & \Rightarrow & \left\{
  \begin{array}{l}
  |v_{1,1}\rangle=|100\rangle \\
  |v_{1,2}\rangle=|101\rangle
  \end{array}\right. \\
 \lambda_2 = 1\ (\times 3) & \Rightarrow & \left\{
  \begin{array}{l}
  |v_{2,1}\rangle=|010\rangle \\
  |v_{2,2}\rangle=|011\rangle \\
  |v_{2,3}\rangle=|110\rangle
  \end{array}\right. \\
 \lambda_3 = 2 & \Rightarrow & |v_3\rangle=|000\rangle-|001\rangle \\
 \lambda_4 \approx 1.438 & \Rightarrow & |v_4\rangle = |000\rangle + |001\rangle - 0.562 |111\rangle \\
 \lambda_5 \approx 5.562 & \Rightarrow & |v_5\rangle = |000\rangle + |001\rangle + 3.562 |111\rangle
 \end{array}\right. ,
 \nonumber
\ee
which contains three entangled eigenvectors and, therefore, ensures
tripartite entanglement for the $abd$ system.

The same trend is found when we proceed with the partial trace over $b$, which
renders the reduced density matrix
\be
 \hat{\rho}_{acd} = \left(\begin{array}{cccc|cccc}
 3 & 0 & 0 & 0 & 0 & 0 & 0 & 1 \\
 0 & 1 & 0 & 0 & 0 & 0 & 0 & 0 \\
 0 & 0 & 0 & 0 & 0 & 0 & 0 & 0 \\
 0 & 0 & 0 & 2 & 0 & 0 & 0 & 0 \\ \hline
 0 & 0 & 0 & 0 & 1 & 0 & 0 & 1 \\
 0 & 0 & 0 & 0 & 0 & 1 & 0 & 0 \\
 0 & 0 & 0 & 0 & 0 & 0 & 0 & 0 \\
 1 & 0 & 0 & 0 & 1 & 0 & 0 & 4
 \end{array} \right) ,
 \label{matrix4}
\ee
with eigenvalues for its transposes being
\be
 \hat{\tilde{\rho}}_{acd}^{T_a}:\ \left\{\begin{array}{l}
 \lambda_1 =0\ (\times 2) \\
 \lambda_2 =1\ (\times 2) \\
 \lambda_3 =3 \\
 \lambda_4 \approx 0.186 \\
 \lambda_5 \approx 2.471 \\
 \lambda_6 \approx 4.343
 \end{array}\right. , \qquad
 \hat{\tilde{\rho}}_{acd}^{T_c} ,\ \hat{\tilde{\rho}}_{acd}^{T_d}:\ \left\{\begin{array}{l}
 \lambda_1 = \phantom{-}0 \\
 \lambda_2 = -1 \\
 \lambda_3 = \phantom{-}1\ (\times 2) \\
 \lambda_4 = \phantom{-}2\ (\times 2) \\
 \lambda_5 = \phantom{-}3 \\
 \lambda_6 = \phantom{-}4
 \end{array}\right. ,
%
 \nonumber
\ee
and eigenvectors
\be
 \left\{\begin{array}{lcl}
 \lambda_1=0\ (\times 2) & \Rightarrow & \left\{
  \begin{array}{l}
  |v_{1,1}\rangle=|010\rangle \\
  |v_{1,2}\rangle=|110\rangle
  \end{array}\right. \\
 \lambda_2=1\:(\times\,2) & \Rightarrow & \left\{
  \begin{array}{l}
  |v_{2,1}\rangle=|001\rangle \\
  |v_{2,2}\rangle=|101\rangle
  \end{array}\right. \\
 \lambda_3=2 & \Rightarrow &
 |v_3\rangle=|011\rangle \\
 \lambda_4\approx 0.657 & \Rightarrow & |v_4\rangle = -0.427|000\rangle-2.916|100\rangle+|111\rangle \\
 \lambda_5\approx 2.529 & \Rightarrow & |v_5\rangle = -2.125|000\rangle+0.654|100\rangle+|111\rangle \\
 \lambda_6\approx 4.814 & \Rightarrow & |v_6\rangle = \phantom{-}0.551|000\rangle-0.262|100\rangle+|111\rangle
 \end{array}\right. ,
 \nonumber
\ee
or when we trace over the qubit $a$ to obtain the reduced matrix
\be
 \hat{\tilde{\rho}}_{bcd} = \left(\begin{array}{cccc|cccc}
 2 & 0 & 0 & 0 & 0 & 0 & 0 & 1 \\
 0 & 1 & 0 & 0 & 0 & 0 & 0 & 0 \\
 0 & 0 & 0 & 0 & 0 & 0 & 0 & 0 \\
 0 & 0 & 0 & 1 & 0 & 0 & 0 & 0 \\ \hline
 0 & 0 & 0 & 0 & 2 & 0 & 0 & 1 \\
 0 & 0 & 0 & 0 & 0 & 1 & 0 & 0 \\
 0 & 0 & 0 & 0 & 0 & 0 & 0 & 0 \\
 1 & 0 & 0 & 0 & 1 & 0 & 0 & 5
 \end{array} \right) ,
 \label{matrix5}
\ee
with eigenvalues for its transposes being
\be
 \hat{\tilde{\rho}}_{bcd}^{T_b}:\;\frac1{12}\left\{
 \begin{array}{l}
 \lambda_1 = 0\ (\times 2) \\
 \lambda_2 = 1\ (\times 2) \\
 \lambda_3 = 2 \\
 \lambda_4 \approx 0.319 \\
 \lambda_5 \approx 2.358 \\
 \lambda_6 \approx 5.323
 \end{array}\right. , \quad
 \hat{\tilde{\rho}}_{bcd}^{T_c} ,\ \hat{\tilde{\rho}}_{bcd}^{T_d}:\;\frac1{12}\left\{
 \begin{array}{l}
 \lambda_1 = \phantom{-}0 \\
 \lambda_2 = -1 \\
 \lambda_3 = \phantom{-}1\ (\times 2) \\
 \lambda_4 = \phantom{-}2\ (\times 3) \\
 \lambda_5 = \phantom{-}5
 \end{array}\right. ,
%
 \nonumber
\ee
and with eigenvectors
\be
 \left\{\begin{array}{lcl}
 \lambda_1=0\:(\times\,2) & \Rightarrow & \left\{
  \begin{array}{l}
  |v_{1,1}\rangle=|010\rangle \\
  |v_{1,2}\rangle=|110\rangle
  \end{array}\right. \\
 \lambda_2=1\:(\times\,3) & \Rightarrow & \left\{
  \begin{array}{l}
  |v_{2,1}\rangle=|001\rangle \\
  |v_{2,2}\rangle=|011\rangle \\
  |v_{2,3}\rangle=|101\rangle
  \end{array}\right. \\
 \lambda_3=2 & \Rightarrow & |v_3\rangle=|000\rangle-|100\rangle \\
 \lambda_4 = 1.438 & \Rightarrow & |v_4\rangle = |000\rangle + |100\rangle - 0.562 |111\rangle \\
 \lambda_5 = 5.562 & \Rightarrow & |v_5\rangle = |000\rangle + |100\rangle + 3.562 |111\rangle
 \end{array}\right. .
 \nonumber
\ee
So, in these two latter cases, again we find a certain amount of tripartite
entanglement for the systems $acd$ and $bcd$, respectively.
In sum, from this analysis, we can conclude that, even though it is weak, there
is certain amount of tripartite entanglement and, therefore, the corresponding
polynomial should include the terms $abc$, $abd$, $acd$ and $bcd$.

The above analysis referred to tripartite entanglement.
We can still descent a level and check whether there is bipartite entanglement
and, if so, to determine which parties are involved.
Accordingly, we are going to compute the doubly reduced matrices (i.e., tracing
the above matrices over another qubit) and check the positivity of the eigenvalues
of the associated partially transpose matrices.
Proceeding this way, we find that
\be
 \hat{\tilde{\rho}}_{ab} = \hat{\tilde{\rho}}_{cd} = \left(\begin{array}{cccc}
 4 & 0 & 0 & 1 \\
 0 & 2 & 0 & 0 \\
 0 & 0 & 0 & 0 \\
 1 & 0 & 0 & 6 \\
 \end{array} \right) , \qquad {\rm with} \quad
 \hat{\tilde{\rho}}_{ab}^{T_a},\ \hat{\tilde{\rho}}_{cd}^{T_c}:\ \left\{\begin{array}{l}
 \lambda_1 = 4 \\
 \lambda_2 = 6 \\
 \lambda_3 \approx -0.414 \\
 \lambda_4 \approx  2.414
 \end{array}\right. ,
 \label{matrix6}
\ee
\be
 \hat{\tilde{\rho}}_{ad} = \hat{\tilde{\rho}}_{bc} = \left(\begin{array}{cccc}
 3 & 0 & 0 & 0 \\
 0 & 3 & 0 & 0 \\
 0 & 0 & 1 & 0 \\
 0 & 0 & 0 & 5 \\
 \end{array} \right), \qquad {\rm with} \quad
 \hat{\tilde{\rho}}_{ad}^{T_a},\ \hat{\tilde{\rho}}_{bc}^{T_b}:\ \left\{\begin{array}{l}
 \lambda_1=1 \\
 \lambda_2=3\ (\times 2) \\
 \lambda_3=5
 \end{array}\right. ,
 \label{matrix8}
\ee
\be
 \hat{\tilde{\rho}}_{ac} = \left( \begin{array}{cccc}
 4 & 0 & 0 & 0 \\
 0 & 2 & 0 & 0 \\
 0 & 0 & 2 & 0 \\
 0 & 0 & 0 & 4 \\
 \end{array} \right), \qquad {\rm with} \quad
 \hat{\tilde{\rho}}_{ac}^{T_a}:\ \left\{\begin{array}{l}
 \lambda_1=2\ (\times 2) \\
 \lambda_2=4\ (\times 2)
 \end{array}\right. ,
 \label{matrix7}
\ee
%
%
and
\be
 \hat{\tilde{\rho}}_{bd} = \left(\begin{array}{cccc}
 2 & 0 & 0 & 0 \\
 0 & 2 & 0 & 0 \\
 0 & 0 & 2 & 0 \\
 0 & 0 & 0 & 6 \\
 \end{array} \right) , \qquad {\rm with} \quad
 \hat{\tilde{\rho}}_{bd}^{T_b}:\ \left\{\begin{array}{l}
 \lambda_1=2\ (\times 3) \\
 \lambda_2=6
 \end{array}\right. .
 \label{matrix10}
\ee
%
%
%
From these results, we noticed that only the reduced systems $ab$ and
$cd$ have negative eigenvalues.
Therefore, only these systems will present entanglement.

From the whole analysis here, we thus conclude that the polynomial
corresponding to the quantum state $|\Psi^{34}\rangle$ is going to
contain six terms, namely four terms accounting for tripartite
entanglement relations and two for bipartite ones.
The final polynomial thus reads as
\be
 \mathcal{P} (\Psi^{34}) = abc + abd + acd + bcd + ab + cd .
\ee


\section{Ring link structures and graphs for $N = 2, 3$ and $4$}
\label{appB}

\begin{figure}[p]
	\begin{center}
		\includegraphics[width=1.00\textwidth]{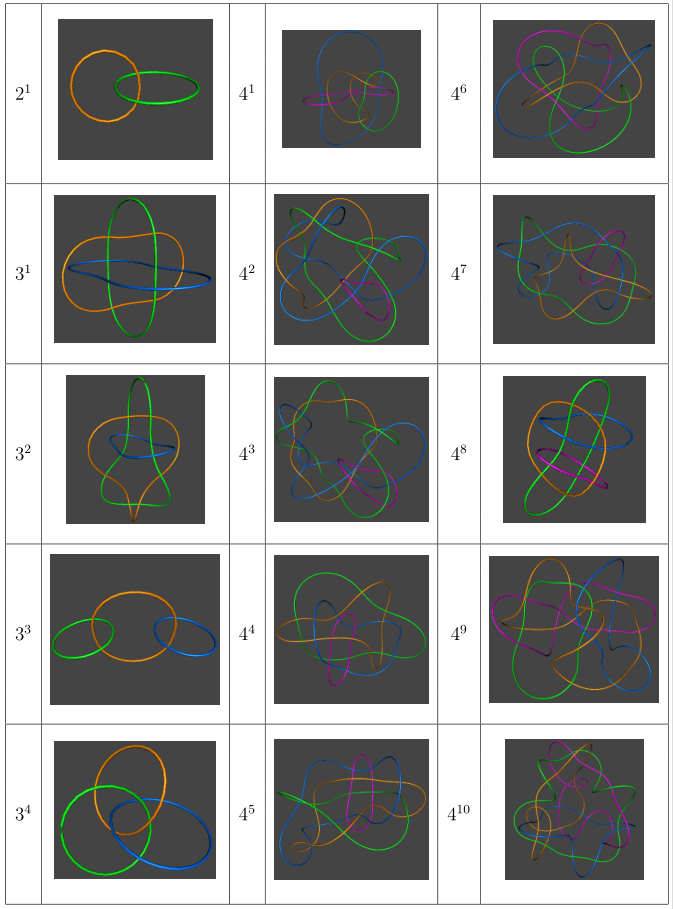}
		\caption{Ring link structures associated with classes $2^1$ to $4^{10}$.}
		\label{FigA1}
	\end{center}
\end{figure}

\begin{figure}[p]
	\begin{center}
		\includegraphics[width=1.00\textwidth]{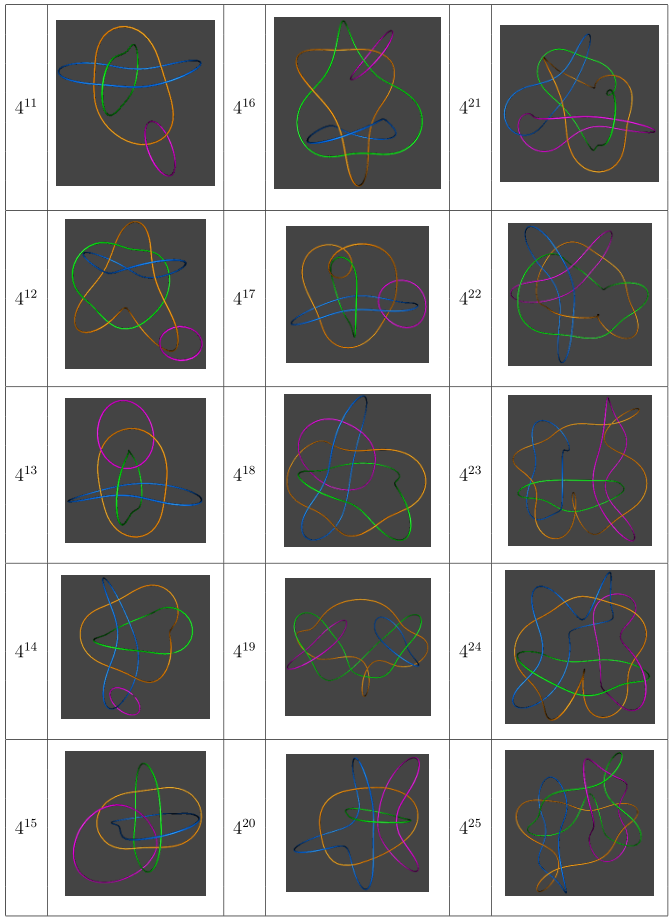}
		\caption{Ring link structures associated with classes $4^{11}$ to $4^{25}$.}
		\label{FigA2}
	\end{center}
\end{figure}

\begin{figure}[p]
	\begin{center}
		\includegraphics[width=1.00\textwidth]{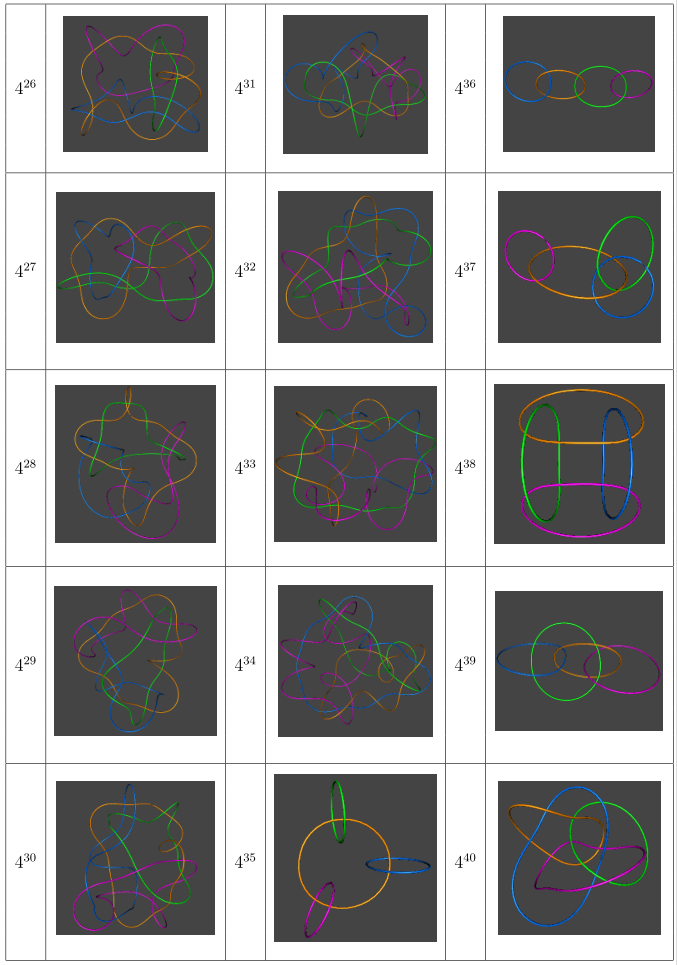}
		\caption{Ring link structures associated with classes $4^{26}$ to $4^{40}$.}
		\label{FigA3}
	\end{center}
\end{figure}

\begin{figure}[p]
	\begin{center}
		\includegraphics[width=1.00\textwidth]{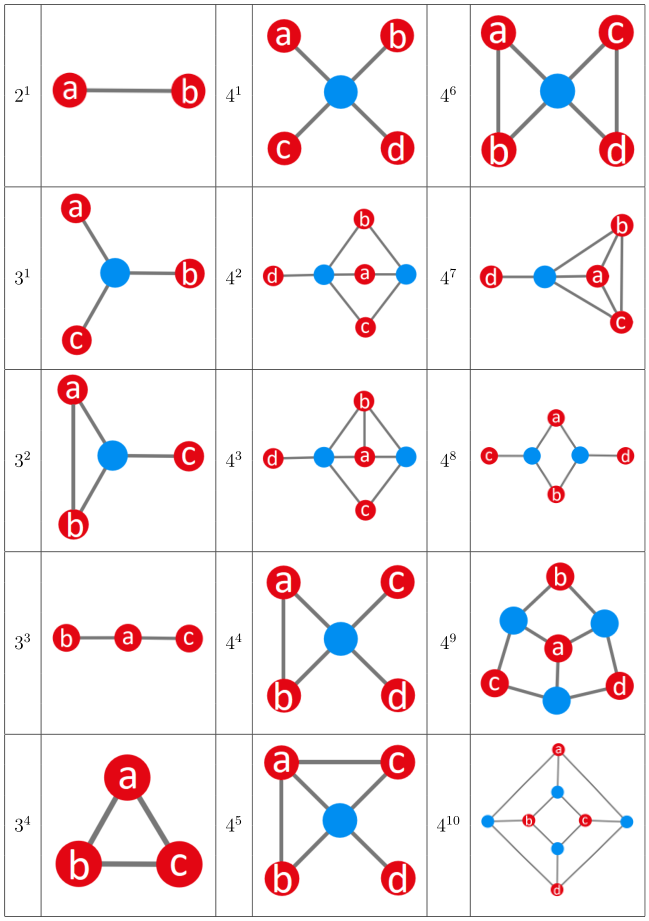}
		\caption{Graphs associated with classes $2^1$ to $4^{10}$.}
		\label{FigA4}
	\end{center}
\end{figure}

\begin{figure}[p]
	\begin{center}
		\includegraphics[width=1.00\textwidth]{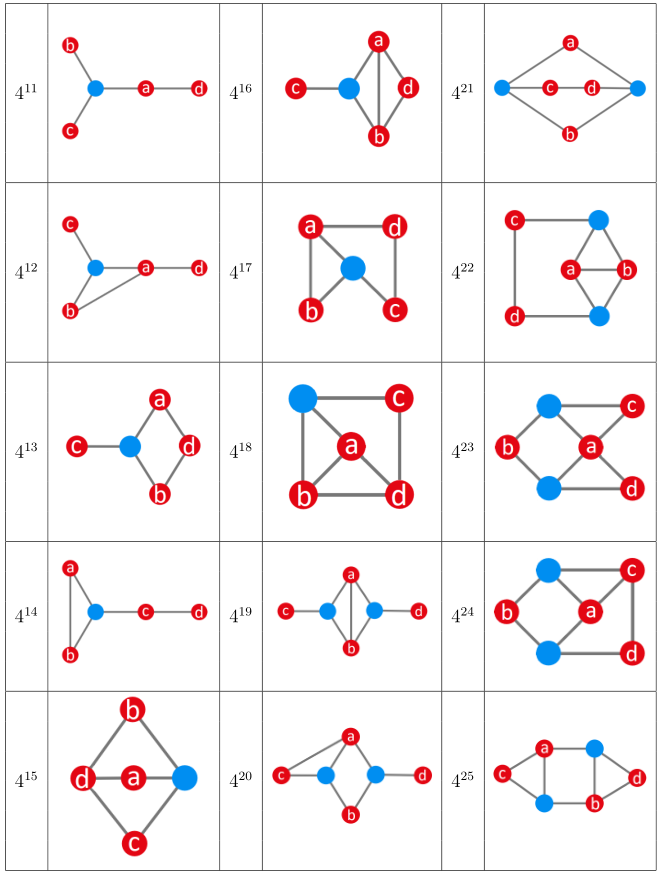}
		\caption{Graphs associated with classes $4^{11}$ to $4^{25}$.}
		\label{FigA5}
	\end{center}
\end{figure}

\begin{figure}[p]
	\begin{center}
		\includegraphics[width=1.00\textwidth]{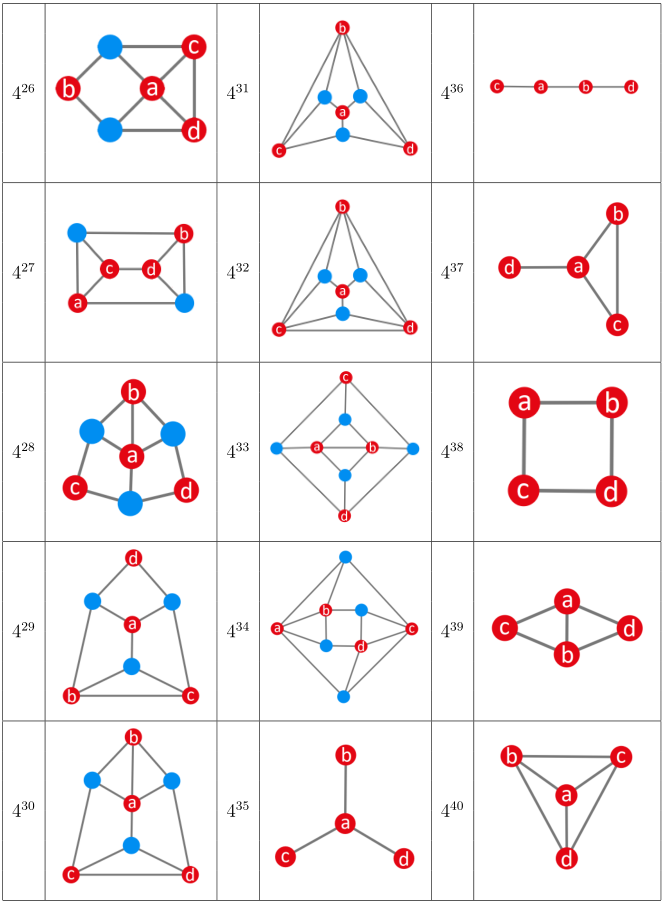}
		\caption{Graphs associated with classes $4^{26}$ to $4^{40}$.}
		\label{FigA6}
	\end{center}
\end{figure}

To be self-contained, and completing the information provided by Quinta and Andr\'e \cite{quinta:PRA:2018},
in this appendix we provide the $N$-ring link structures (Figs.~\ref{FigA1}, \ref{FigA2} and \ref{FigA3}) and
graphs (Figs.~\ref{FigA4}, \ref{FigA5} and \ref{FigA6}) for all the possible classes for $N=2, 3$ and 4
(the correspondence between links and polynomials for $N=4$ can be found in \cite{quinta:PRA:2018}).




\reftitle{References}

\externalbibliography{no}

\begin{thebibliography}{-------}
\providecommand{\natexlab}[1]{#1}

\bibitem[Dowling and Milburn(2003)]{milburn:PTRSLA:2003}
Dowling, J.P.; Milburn, G.J.
\newblock Quantum technology: The second quantum revolution.
\newblock {\em Phil. Trans. Roy. Soc. Lond. A} {\bf 2003}, {\em
  361},~1655--1674.

\bibitem[Gisin(2019)]{gisin:entropy:2019}
Gisin, N.
\newblock Entanglement 25 years after quantum teleportation: testing joint
  measurements in quantum networks.
\newblock {\em Entropy} {\bf 2019}, {\em 21},~325(1--12).

\bibitem[Bennett \em{et~al.}(1993)Bennett, Brassard, Cr\'epeau, Jozsa, Peres,
  and Wootters]{bennet:PRL:1993}
Bennett, C.H.; Brassard, G.; Cr\'epeau, C.; Jozsa, R.; Peres, A.; Wootters,
  W.K.
\newblock Teleporting an unknown quantum state via dual classical and
  Einstein-Podolsky-Rosen channels.
\newblock {\em Phys. Rev. Lett.} {\bf 1993}, {\em 70},~1895--1899.

\bibitem[Bouwmeester \em{et~al.}(1997)Bouwmeester, Pan, Mattle, Eibl,
  Weinfurter, and Zeilinger]{bouwmeester:nature:1997}
Bouwmeester, D.; Pan, J.W.; Mattle, K.; Eibl, M.; Weinfurter, H.; Zeilinger, A.
\newblock Experimental quantum teleportation.
\newblock {\em Nature} {\bf 1997}, {\em 390},~575--579.

\bibitem[Marcikic \em{et~al.}(2003)Marcikic, de~Riedmatten, Tittel, Zbinden,
  and Gisin]{zeilinger:nature:2003}
Marcikic, I.; de~Riedmatten, H.; Tittel, W.; Zbinden, H.; Gisin, N.
\newblock Long-distance teleportation of qubits at telecommunication
  wavelengths.
\newblock {\em Nature} {\bf 2003}, {\em 421},~509--513.

\bibitem[Ma \em{et~al.}(2012)Ma, Herbst, Scheidl, Wang, Kropatschek, Naylor,
  Wittmann, Mech, Kofler, Anisimova, Makarov, Jennewein, Ursin, and
  Zeilinger]{zeilinger:nature:2012}
Ma, X.S.; Herbst, T.; Scheidl, T.; Wang, D.; Kropatschek, S.; Naylor, W.;
  Wittmann, B.; Mech, A.; Kofler, J.; Anisimova, E.; Makarov, V.; Jennewein,
  T.; Ursin, R.; Zeilinger, A.
\newblock Quantum teleportation over 143 kilometres using active feed-forward.
\newblock {\em Nature} {\bf 2012}, {\em 489},~269--273.

\bibitem[Yin \em{et~al.}(2017{\natexlab{a}})Yin, Cao, Li, Liao, Zhang, Ren,
  Cai, Liu, Li, Dai, Li, Lu, Gong, Xu, Li, Li, Yin, Jiang, Li, Jia, Ren, He,
  Zhou, Zhang, Wang, Chang, Zhu, Liu, Chen, Lu, Shu, Peng, Wang, and
  Pan]{JWPan:science:2017}
Yin, J.; Cao, Y.; Li, Y.H.; Liao, S.K.; Zhang, L.; Ren, J.G.; Cai, W.Q.; Liu,
  W.Y.; Li, B.; Dai, H.; Li, G.B.; Lu, Q.M.; Gong, Y.H.; Xu, Y.; Li, S.L.; Li,
  F.Z.; Yin, Y.Y.; Jiang, Z.Q.; Li, M.; Jia, J.J.; Ren, G.; He, D.; Zhou, Y.L.;
  Zhang, X.X.; Wang, N.; Chang, X.; Zhu, Z.C.; Liu, N.L.; Chen, Y.A.; Lu, C.Y.;
  Shu, R.; Peng, C.Z.; Wang, J.Y.; Pan, J.W.
\newblock Satellite-based entanglement distribution over 1200 kilometers.
\newblock {\em Science} {\bf 2017}, {\em 356},~1140--1144.

\bibitem[Yin \em{et~al.}(2017{\natexlab{b}})Yin, Cao, Li, Ren, Liao, Zhang,
  Cai, Liu, Li, Dai, Li, Huang, Deng, Li, Zhang, Liu, Chen, Lu, Shu, Peng,
  Wang, and Pan]{JWPan:PRL:2017}
Yin, J.; Cao, Y.; Li, Y.H.; Ren, J.G.; Liao, S.K.; Zhang, L.; Cai, W.Q.; Liu,
  W.Y.; Li, B.; Dai, H.; Li, M.; Huang, Y.M.; Deng, L.; Li, L.; Zhang, Q.; Liu,
  N.L.; Chen, Y.A.; Lu, C.Y.; Shu, R.; Peng, C.Z.; Wang, J.Y.; Pan, J.W.
\newblock Satellite-to-ground entanglement-based quantum key distribution.
\newblock {\em Phys. Rev. Lett.} {\bf 2017}, {\em 119},~200501(1--5).

\bibitem[Li and Yin(2016)]{li:SciBull:2016}
Li, T.C.; Yin, Z.Q.
\newblock Quantum superposition, entanglement, and state teleportation of a
  microorganism on an electromechanical oscillator.
\newblock {\em Sci. Bull.} {\bf 2016}, {\em 61},~163--171.

\bibitem[Landsman \em{et~al.}(2019)Landsman, Figgatt, Schuster, Linke, Yoshida,
  Yao, and Monroe]{landsman:nature:2019}
Landsman, K.A.; Figgatt, C.; Schuster, T.; Linke, N.M.; Yoshida, B.; Yao, N.Y.;
  Monroe, C.
\newblock Verified quantum information scrambling.
\newblock {\em Nature} {\bf 2019}, {\em 567},~61--64.

\bibitem[Luo \em{et~al.}(2019)Luo, Zhong, Erhard, Wang, Peng, Krenn, Jiang, Li,
  Liu, Lu, Zeilinger, and Pan]{zelinger-JWPan:PRL:2019}
Luo, Y.H.; Zhong, H.S.; Erhard, M.; Wang, X.L.; Peng, L.C.; Krenn, M.; Jiang,
  X.; Li, L.; Liu, N.L.; Lu, C.Y.; Zeilinger, A.; Pan, J.W.
\newblock Quantum teleportation in high dimensions.
\newblock {\em Phys. Rev. Lett.} {\bf 2019}, {\em 123},~070505(1--4).

\bibitem[Penrose(1971)]{penrose:graphnot:1971}
Penrose, R., Combinatorial Mathematics and its Applications; Academic Press:
  New York,  1971; chapter Applications of negative dimensional tensors, pp.
  221--244.

\bibitem[Coecke(2010)]{coecke:contempphys:2010}
Coecke, B.
\newblock Quantum picturalism.
\newblock {\em Contemp. Phys.} {\bf 2010}, {\em 51},~59--83.

\bibitem[Kauffman and Lomonaco(2002)]{kauffman:NJP:2002}
Kauffman, L.H.; Lomonaco, S.J.
\newblock Quantum entanglement and topological entanglement.
\newblock {\em New J. Phys.} {\bf 2002}, {\em 4},~73(1--18).

\bibitem[Kauffman and Lomonaco(2004)]{kauffman:NJP:2004}
Kauffman, L.H.; Lomonaco, S.J.
\newblock Braiding operators are universal quantum gates.
\newblock {\em New J. Phys.} {\bf 2004}, {\em 6},~134(1--40).

\bibitem[Mironov(2019)]{mironov:universe:2019}
Mironov, S.
\newblock Topological entanglement and knots.
\newblock {\em Universe} {\bf 2019}, {\em 5},~60(1--10).

\bibitem[Melnikov \em{et~al.}(2019)Melnikov, Mironov, Mironov, Morozov, and
  Morozov]{mironov:JHEP:2019}
Melnikov, D.; Mironov, A.; Mironov, S.; Morozov, A.; Morozov, A.
\newblock From topological to quantum entanglement.
\newblock {\em J. High Ener. Phys.} {\bf 2019}, {\em 2019},~116(1--12).

\bibitem[Karlsson and Bourennane(1998)]{karlsson:PRA:1998}
Karlsson, A.; Bourennane, M.
\newblock Quantum teleportation using three-particle entanglement.
\newblock {\em Phys. Rev. A} {\bf 1998}, {\em 58},~4394--4400.

\bibitem[Quinta and Andr\'e(2018)]{quinta:PRA:2018}
Quinta, G.M.; Andr\'e, R.
\newblock Classifying quantum entanglement through topological links.
\newblock {\em Phys. Rev. A} {\bf 2018}, {\em 97},~042307(1--12).

\bibitem[Peres(1996)]{peres:PRL:1996}
Peres, A.
\newblock Separability criterion for density matrices.
\newblock {\em Phys. Rev. Lett.} {\bf 1996}, {\em 77},~1413--1415.

\bibitem[Horodecki \em{et~al.}(1996)Horodecki, Horodecki, and
  Horodecki]{horodecki:PLA:1996}
Horodecki, M.; Horodecki, P.; Horodecki, R.
\newblock Separability of mixed states: necessary and sufficient conditions.
\newblock {\em Phys. Lett. A} {\bf 1996}, {\em 223},~1--8.

\bibitem[D\"ur \em{et~al.}(1999)D\"ur, Cirac, and Tarrach]{cirac:PRL:1999}
D\"ur, W.; Cirac, J.I.; Tarrach, R.
\newblock Separability and distillability of multiparticle quantum systems.
\newblock {\em Phys. Rev. Lett.} {\bf 1999}, {\em 83},~3562--3565.

\bibitem[D\"ur and Cirac(2000)]{cirac:PRA:2000}
D\"ur, W.; Cirac, J.I.
\newblock Classification of multiqubit mixed states: Separability and
  distillability properties.
\newblock {\em Phys. Rev. A} {\bf 2000}, {\em 61},~042314(1--11).

\end{thebibliography}


\end{document}